\documentclass{aa}  
\usepackage{graphicx}
\usepackage{txfonts}
\begin{document}
\title{The cycle of interstellar dust in galaxies of different morphological types}

\author{Francesco Calura$^{\,1}$, Antonio Pipino$^{\,2,3}$
Francesca Matteucci$^{\,1,2}$}
\institute{1 INAF - Osservatorio Astronomico di Trieste, via G. B. Tiepolo 11, 34131 Trieste, Italy \\
2 Dipartimento di Astronomia - Universit\'a di Trieste, Via G. B. Tiepolo 11, 34131 Trieste, Italy \\
3 Astrophysics, University of Oxford, Denys Wilkinson Building, Keble Road, Oxford OX1 3RH, U.K.}

\offprints{F. Calura, \email{fcalura@oats.inaf.it} }

\date{Accepted}

\abstract
{By means of chemical evolution models for galaxies of different morphological type, 
we have performed a detailed study of the evolution of the cosmic dust properties in different environments: 
the solar neighbourhood, elliptical galaxies and dwarf irregular galaxies.}
{Starting from the same formalism as developed 
by Dwek (1998), and thanks to the uptodate observations available in the solar vicinity, we intend to study the effects of dust in the chemical evolution of different types of galaxies and at the same time to  refine the investigation of 
the parameter space and achieving a satisfactory  fine tuning of the parameters involved in our study.} 
{We have taken into account dust production from low and intermediate mass stars, supernovae II and Ia as well as dust destruction and dust accretion processes in a detailed model of chemical evolution  for the solar vicinity. Then, by means of the same dust prescriptions but adopting different galactic models (different star formation histories and presence of galactic winds), 
we have extended our study to ellipticals and dwarf irregular galaxies. 
In all these systems, dust evolution has been calculated by means of chemical evolution models which
relax the instantaneous recycling 
approximation and already reproduce the main features of the various galaxies.}
{We have investigated how the assumption of different star formation histories 
affects the dust production rates, the dust depletion, the dust accretion and destruction rates.
We have predicted dust to gas and dust to metals ratios in very good agreement with those observed in the solar vicinity. 
We have shown how the inclusion of the dust treatment is helpful in solving the so-called Fe discrepancy, observed in the hot gaseous halos of local ellipticals, and in reproducing the chemical abundances observed in the Lyman Break Galaxies. Finally, our new models can be very useful in future detailed spectro-photometric studies of galaxies. }{}
\keywords{ISM: dust, extinction; ISM: abundances; galaxies: abundances; galaxies: evolution. }

\authorrunning{Calura, Pipino, Matteucci}
\maketitle
\section{Introduction}
The presence of dust in local and distant galaxies is indicated 
by various observational evidences. 
The light emitted by stars interacts with the dust 
grains in a wavelength-dependent manner: the light 
emitted in the ultra-violet and optical bands is absorbed and scattered 
by the dust grains. This is the well-known phenomenon called 
dust extinction, which is taking place both inside and outside galaxies (Aguirre 1999). 
In the optical band, the larger is the emission 
wavelength, the lower is the extinction effect caused by dust.\\ 
In our Galaxy and in local starburst galaxies, 
dust grains are the main contributors to the 
emission in the mid and far infrared bands. In fact, 
the light absorbed by dust at UV and optical wavelengths is then 
thermally re-emitted at much longer wavelengths, 
in the range 10-1000 $\mu$m. 
Also  the chemical composition of the galactic interstellar medium 
(ISM) is strongly influenced by the presence of dust. 
Some chemical elements, called refractory, in the gas phase are 
subject to dust depletion and a fraction of their total abundance 
is incorporated into solid  grains. 
Examples of refractory elements  are 
Fe, Si, Mg, Ni. 
For these elements, the abundance measured in stellar 
surfaces is considerably higher than the gas phase 
abundances (Li 2005).\\ 
 Dwek (1998, hereinafter D98) has developed a chemical evolution model to study the dust content 
of the Milky Way galaxy and its evolution. 
He has focused on some refractory chemical species and for each one    
he has suggested a set of dust condensation efficiencies, calculating the dust production rates  
from low and intermediate mass stars, type Ia and type II supernovae (SNe). For these elements, he has  
investigated also the dust destruction and the accretion rates, 
the dust to gas ratios and the dust fractions at the present time. 

Since then, the amount of observational data regarding the local 
content of dust has noticeably grown (Draine 2003), 
allowing a reliable fine tuning of the parameters involved in theoretical dust evolution studies. 
By means of a one-zone chemical evolution for the solar neighbourhood, Zhukovska, Gail \& Trieloff (2007)
studied production by stars of various dust species. 
By considering the available observations of presolar dust grains in meteorites, their 
study  has allowed them to put some constraints on the dust condensation efficiencies from SNe and AGB stars. 
Moving to extragalactic objects, new observations call for a theoretical investigation of 
the dust content of galaxies of different morphological types. In particular, the SCUBA camera has revealed
a large number of dusty sources at high redshfit which might be young massive ellipticals undergoing a 
starburst (Lilly et al. 1999, Eales et al. 2000). For what concerns, instead, local ellipticals, new measurements have rejuvenated the interest
in assessing the issue of the so-called \emph{iron-discrepancy} (Arimoto et al. 1997) by means of the dust.  
In this paper, we present new set of chemical evolution models, with updated nucleosynthesis prescriptions. 
All of these models are successful in reproducing the abundances of various galaxies (spirals, ellipticals, irregulars).  
By means of our chemical evolution models, 
we study dust evolution in various environments. Starting from the same formalism developed by D98, 
first we focus on the Solar Neighbourhood (S.N.) and we extend the analysis by D98, 
improving and deepening the investigation of the parameter space. 
We focus on the differential roles 
of type Ia and II SNe in dust destruction and production and  
the evolution of the accretion and destruction rates. When possible, we compare these quantities 
with the data available in the literature computed by other authors. 
In this paper, we perform 
a chemical evolution study of various dust species in environments different than the Milky Way Galaxy. 
This is possible by taking into account 
in detail the stellar lifetimes, allowing us to predict the properties of dust species such as C and Fe, produced by stars with 
lifetimes spanning from $\sim 0.03$ Gyr up to $\sim$  10 Gyr. 
We focus on the effects that different star formation histories 
have on the evolution of the dust content of galaxies of different morphological types.\\
This paper is organized as follows:   
in section 2 we present the chemical evolution equations used to compute dust evolution and all the parameters involved. 
In section 3 we discuss our results for the S.N., for elliptical galaxies and for dwarf 
irregular galaxies. Finally, in Section 4 we draw some conclusions.

\section{The chemical evolution model}  

\subsection{The chemical evolution equations}
By means of detailed chemical evolution models,  
it is possible to follow the evolution of the abundances of several chemical
species and of the dust content of spirals, elliptical and irregular galaxies. 
In all models the instantaneous recycling approximation is relaxed and the 
stellar lifetimes are taken into account. 
Detailed descriptions of the chemical evolution models can be found in Matteucci
\& Tornamb\'{e} (1987) and Matteucci (1994) for elliptical galaxies, 
Chiappini et al. (1997, 2001) for the spirals and 
Bradamante et al. (1998) for irregular galaxies.\\
In our picture, elliptical galaxies form as a result of the rapid collapse of a homogeneous sphere of
primordial gas where intense star formation (SF) is taking place at the same time as the collapse proceeds. 
SF is assumed to halt as the energy of the ISM, heated by stellar winds and supernova (SN) explosions,
exceeds the binding energy of the gas. At this time a galactic wind occurs, sweeping away almost all of  
the residual gas. After the SF has stopped, the galactic wind is maintained by type Ia SNe, and its duration
depends on the balance between this heating source and the gas cooling (we refer the reader to Pipino et al. 2002, 2005).\\ 
For spiral galaxies, the adopted model is calibrated in order to reproduce a large set of observational 
constraints for the Milky Way galaxy (Chiappini et al. 2001). 
The Galactic disc is approximated by several independent rings, 
2 kpc wide, without exchange of matter between them. In our picture, 
spiral galaxies are assumed to form as a result of two main infall episodes.  
During the first episode, the halo and the thick disc are formed.
During the second episode, a slower infall
of external gas forms the thin disc with the gas accumulating faster in the inner than in the outer
region ("inside-out" scenario, Matteucci \& Fran\c cois 1989). The process of disc formation is much longer than the
halo 
and bulge formation, with time scales varying from $\sim2$ Gyr in the inner disc to $\sim7$ Gyr in the solar region
and up to $20$ Gyr in the outer disc (see table 1). In this paper, we are interested in the study of 
dust evolution in the S.N. For this purpose, we focus on a ring located at 8 kpc from the 
Galactic centre, 2 kpc wide. \\ 
Finally, irregular galaxies are assumed to assemble from infall of protogalactic small clouds 
of primordial chemical composition, until masses in the range $\sim 10^{9}M_{\odot}$ are accumulated, 
and to produce stars at a lower rate than spirals. \\

All the models used in this paper consider only one gas phase. For a single-phase gas, 
the chemical evolution equation for a given chemical element \emph{i} takes the following form:
\begin{eqnarray}
{d G_i (t) \over d t}  &= & -\psi (t) X_i (t)\,   \nonumber \\
& & +\int_{M_L}^{M_{B_m}} \psi (t-\tau_m) Q_{\rm mi}(t-\tau_m) \phi (m) dm\,  \nonumber \\
& & + A\int_{M_{B_m}}^{M_{B_M}} \phi (m) \cdot \nonumber \\
& & \left [ \int_{\rm\mu_{\rm min}}^{0.5} f(\mu) Q_{\rm mi}(t-\tau_{\rm m_2}) \psi (t-\tau_{\rm m_2}) d\mu \right ] dm \,  \nonumber \\
& &  +(1-A)\, \int_{\rm M_{\rm B_m}}^{M_{\rm B_M}} \psi (t-\tau_m) Q_{\rm mi}(t-\tau_m) \phi (m) dm\,  \nonumber \\
& &  +\int_{\rm M_{\rm B_M}}^{M_U} \psi (t-\tau_m) Q_{\rm mi}(t-\tau_m) \phi (m) dm\,  \nonumber \\
& & +({d G_i (t) \over d t})_{\rm inf} -({d G_i (t) \over d t})_{\rm out}  	
\label{eq_chem}
\end{eqnarray}
where $G_{i}(t)=M_{g}(t)X_{i}(t)/M_{tot}$ is the gas mass in 
the form of an element $i$ normalized to a total fixed mass 
$M_{tot}$ and $G(t)= M_{g}(t)/M_{tot}$ is the total fractional 
mass of gas present in the galaxy at the time $t$. 
The same quantities can be defined in terms of the surface gas and mass 
densities, especially in spiral galaxies. 
\rm $X_i (t)$ is defined as the abundance by mass (or mass fraction) of the element \emph{i} 
(for a comprehensive discussion of this equation, see Matteucci $\&$ Greggio 1986).
\rm The quantity $\psi (t)$ is the star formation rate (SFR). 
The term $ -\psi (t) X_i (t)$ gives the rate at which
the element \emph{i} is subtracted from the ISM by  the SF process. 
The second term is the rate at which each element is restored into the ISM 
by single stars with masses in the range $M_{L}$ - $M_{B_m}$, where $M_{L}$ is the minimum mass contributing, 
at a given time $t$, to chemical enrichment 
(the minimum is $0.8 M_{\odot}$) 
and $M_{B_m}$ is the minimum binary mass allowed for binary systems giving rise to type Ia SN ($3 M_{\odot}$, Matteucci \& Greggio 1986).  
The quantities $Q_{mi}(t-\tau_m)$ (where $\tau_m$ is the lifetime of a star of mass $m$) contain all
the information about stellar nucleosynthesis for elements either produced or destroyed inside 
stars or both (Talbot and Arnett 1971).  
The third term represents the enrichment due to binaries which become type Ia SN, i.e. all the binary systems with total mass 
between $M_{B_m}$ and $M_{B_M}=16 M_{\odot}$. For the type Ia SN progenitor model, the Single Degenerate (SD) scenario is assumed, 
where a C-O white dwarf explodes by C-deflagration mechanism 
after having reached the Chandrasekhar mass ($1.44 M_{\odot}$), owing to progressive 
mass accretion from a non-degenerate companion (Whelan \& Iben 1973). 
The parameter $A$ represents the unknown fraction of binary stars
giving rise to type Ia SN and is fixed by reproducing the observed 
present time SN Ia rate. 
In this third term, both quantities $\psi$ and $Q_{mi}$ refer to the time $t-t_{m_2}$, where $t_{m_2}$ indicates the lifetime of the 
secondary star of the binary system, which regulates the explosion timescale. 
$\mu=M_{2}/M_{B}$ is the ratio between the mass of the secondary component $M_{2}$ 
and the total mass of the binary system $M_{B}$, whereas 
$f(\mu)$ is the 
distribution function of this ratio. Statistical studies indicate that mass 
ratios close to $0.5$ are preferred, so the formula:\\
\begin{equation}
f(\mu)=2^{1+\gamma}(1+\gamma)\mu^{\gamma}
\end{equation}
is commonly adopted, with $\gamma=2$ (Greggio \& Renzini 1983). 
$\mu_{min}$ is the minimum mass fraction contributing to the SNIa rate at the time $t$, and is given by:\\ 
\begin{equation}
\mu_{min}=max \left \{ \frac{M_{2}(t)}{M_{B}}, \frac{M_{2}-0.5M_{B}}{M_{B}} \right \}
\end{equation}
The fourth term represents the enrichment due to stars in the mass range $M_{B_m}$ - $M_{B_M}$ which are either single, or, 
if in binaries, do not produce a SN Ia event. In this mass range, all the stars with masses $m > 8 M_{\odot}$ will explode as 
type II SNe, which in our picture are assumed to originate from core collapse of single massive stars. 
The fifth term represents the enrichment of stars more massive than $M_{B_M}$, all of which explode as 
core collapse SNe. 
As the upper mass limits contributing to chemical enrichment, we assume $M_{U}=100 M_{\odot}$. 
Finally, the last two terms account for infall  
of external gas and for galactic winds,
respectively. For the infall term, an exponential law with different timescales 
is adopted for spirals and irregulars (see Calura \& Matteucci 2006b).  
Concerning ellipticals, 
Pipino \& Matteucci (2004) showed that, 
in order to satisfy the largest number of photo-chemical properties, a quick infall, with a shorter timescale for the more
massive objects, is needed. \\
A non-zero outflow term is present in the equations describing ellipticals and irregular galaxies. 
In both of these galaxies, a galactic wind develops as the
thermal energy of the gas heated by SN explosions exceeds the binding
energy of the gas (see Bradamante et al. 1998, Pipino \& Matteucci 2004). 
The binding energy of the gas is strongly 
influenced by assumptions concerning the presence and distribution of dark
matter (Matteucci 1992); for the model adopted here a diffuse 
($R_e/R_d$=0.1, where
$R_e$ is the effective radius of the galaxy and $R_d$ is the radius 
of the dark matter core) but 
massive ($M_{dark}/M_{lum}=10$) dark halo has 
been assumed (see Bertin et al. 1992). In the case of spiral galaxies, the outflow term is set to zero. 

\subsubsection{The initial mass function}

$\phi(m)$ is 
the initial mass function (IMF), assumed to be constant in space and time and normalized to unity in the mass interval $0.1 -100 M_{\odot}$. 
For the spiral galaxies, we adopt a simplified two-slope approximation to the actual  
   Scalo (1986) IMF, similarly to what is done in Matteucci \& Fran\c cois 
   (1989), which is expressed by the formula: 

\begin{displaymath}
     \phi_{\mathrm{Scalo}}(m) = \left\{ \begin{array}{l l}
                                      0.19\, \cdot m^{-1.35} & 
			     \qquad {\mathrm{if}} \; m < 2 \, M_\odot \\
			              0.24\, \cdot m^{-1.70} &
			     \qquad {\mathrm{if}} \; m > 2 \, M_\odot, \\
                                       \end{array} \right.
\end{displaymath}

whereas for ellipticals and irregulars  we adopt a Salpeter (1995) IMF, of the form \\

\begin{math}
\phi_{\mathrm{Salp}} (m) = 0.17 \cdot m^{-1.35}\\
\end{math}

The reason for such a choice relies mainly on the 
chemical abundances and on the metal content observed in each morphological type (see Calura \& Matteucci 2006b). 
In fact, it is well known that a Salpeter (1955) IMF can account for the abundances observed in 
local ellipticals and dwarf galaxies (Pipino \& Matteucci 2004, Recchi et al. 2002), whereas it leads to 
an overestimation of the metal abundances in spiral discs (see Romano et al. 2005) which, on the other hand, 
are well accounted for by means of the  Scalo (1986) IMF.\\

\subsubsection{The nucleosynthesis prescriptions}
The yields used in the present work are separated into three groups: 
yields of low and intermediate mass stars, yields of type Ia SNe and yields of massive stars.\\
Low and intermediate mass stars (i.e. with masses $0.8 M_{\odot}\le m \le  8 M_{\odot}$) 
contribute through quiescent mass loss and planetary nebula phase to the 
ISM metal enrichment. 
type Ia SNe are assumed to originate from exploding white dwarfs in binary systems, characterized by  
total masses $3 M_{\odot} \le m_{bin} \le 16 M_{\odot}$, according to the Matteucci \& Recchi (2001) best model. Finally, 
we assume that single massive stars with initial masses in the range $>8 - 100 M_{\odot}$ explode as core collapse 
SNe.  The nucleosynthesis prescriptions are common to all models. 
For massive stars and type Ia SNe, we adopt the empirical yields suggested by  Fran\c cois et al. (2004),   
which are substantially based on the Woosley \& Weaver (1995) and Iwamoto et al. (1999) yields, respectively, 
and are tuned to reproduce at best the abundances in the S.N.  
For low and intermediate mass stars, we adopt the prescriptions by van den Hoek $\&$ Groenewegen (1997).  \\

\subsubsection{The star formation rate}
\label{Sec_SFR}
For ellipticals and irregulars galaxies, the SFR $\psi(t)$ in our models is a 
Schmidt (1959) law expressed as:

\begin{equation}
\psi(t) = \nu G^{k}(t),  
\end{equation}

with $k=1$. 
Here $G(t)=\sum G_{i}(t)$ is the normalized total gas density or gas fraction at the time $t$.  
The quantity $\nu$ is the efficiency of SF, 
namely the inverse of the typical time-scale for SF,
and is expressed in $Gyr^{-1}$. 

\begin{figure*}
\centering
\vspace{0.001cm}
\includegraphics[height=18pc,width=18pc]{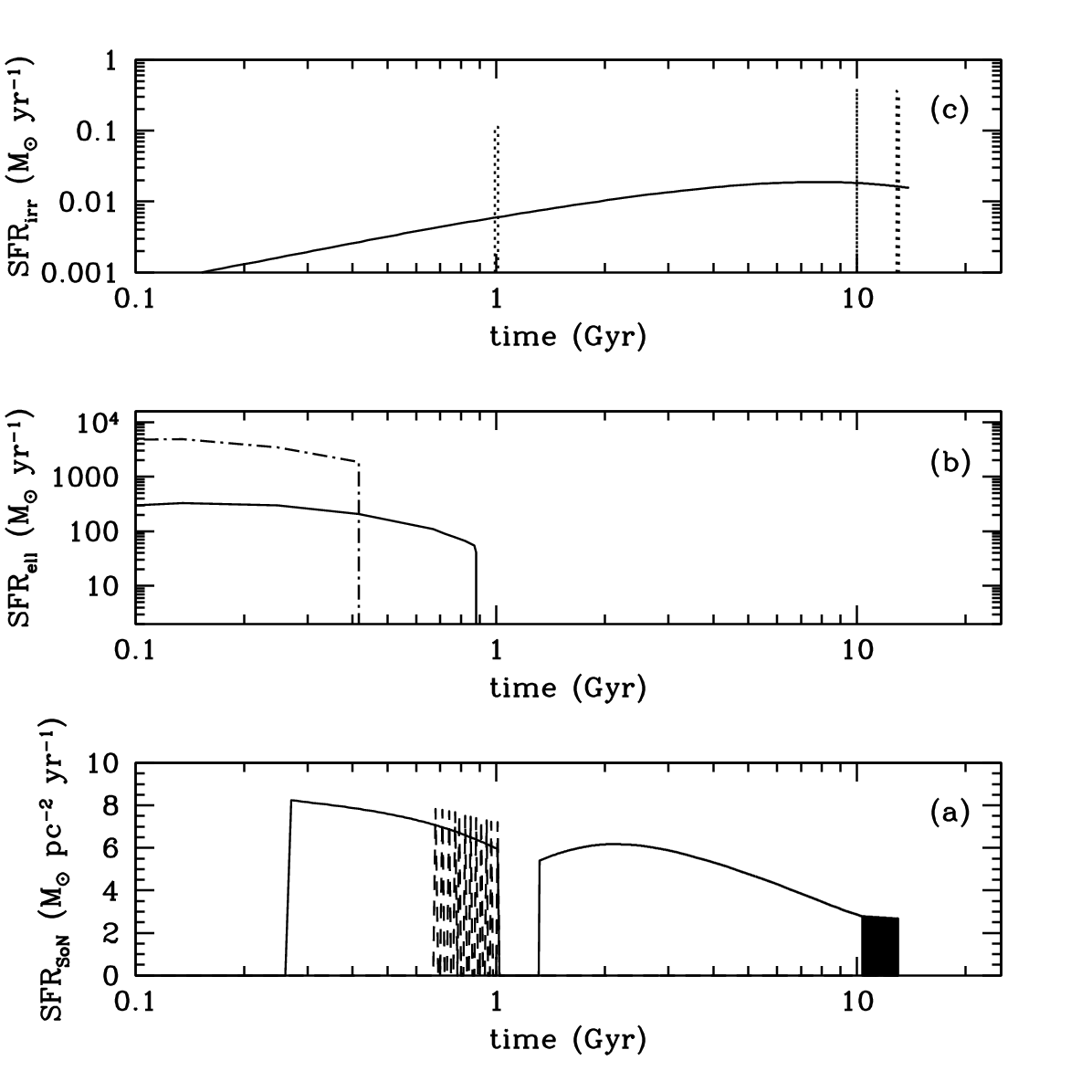}
\caption[]{Predicted SFRs as a function of time for different chemical evolution models. Panel (a): 
S.N. (solid line) and a model for the outer regions of the Milky Way disc (dashed line); 
panel (b): two different elliptical galaxy models, i.e. model La1 (solid line) and Ha1 (dash-dotted line) of Pipino et al. (2005) ; 
panel (c): two different irregulars, one with continuous SF (solid line) and the 
other with 3 starbursts (dotted line). 
}
\label{SFR}
\end{figure*}

In the case of spiral galaxies, the SFR expression (Chiappini et al. 1997) is:

\begin{equation}
\psi(r,t) = \nu [\frac{\sigma(r,t)}{\sigma(r_{\odot},t)}]^{2(k-1)} [\frac{\sigma(r,t_{Gal})}{\sigma(r,t)}]^{k-1}
\sigma^{k}_{ISM}(r,t)
\end{equation}
where $\nu$ is the SF efficiency, $\sigma(r,t)$ is the total mass 
(gas + stars) surface density at a radius r and time t,
$\sigma(r_{\odot},t)$ is the total mass surface density in the solar 
region and $\sigma_{ISM}(r,t)$ is the ISM surface mass
density. For the gas density exponent $k$ a value of 1.5 has been assumed 
by Chiappini et al. (1997) in order to ensure a good
fit to the observational constraints for a large set of local spirals (Kennicutt 1998). 
The efficiency of SF is set to $\nu=1 Gyr^{-1}$, and
becomes zero when the gas surface density drops below a certain 
critical threshold. For the SF, we adopt a threshold gas 
density $\sigma_{th}\sim 7 M_{\odot} pc^{-2}$ in the disc as suggested by 
Kennicutt (1989). 
The difference between equations 4 and 5 is due to the different normalizations of the quantities 
involved in the chemical evolution equations (see section 2.1). Furthermore, the model used for spirals is a multi-zone 
one, where the SFR expression is a function of the galactocentric radius. On the other hand, the models of ellipticals and 
irregulars are one-zone and the SFR has a simpler expression. It is important to note that the parameterization of the SFR 
in both equations 4 and 5 involves two different quantities, i.e. $\nu$ and $k$, which act in the same way. \\
In table 1, we show the adopted parameters for all the  chemical evolution models described in this section.  
In Figure~\ref{SFR}, we show different SFRs for various models used in this work. 
In panel (a), we show the SFRs for the two models considered here for spiral galaxies: 
the S.N. model (solid line) and a model for the disk at a distance  of 16 Kpc  from
the Galactic center (dashed line). Concerning the S. N. model, the most striking features are the 
SF hiatus at 1 Gyr (see Chiappini et al. 1997, 2001) and the threshold-dominated SF after 10 Gyr. 
Both features have important consequences on the behaviour 
of several quantities studied in this paper. 
The model used for ellipticals is the one by Pipino et al. (2005).
This is a first step in the self-consistent study of both optical and X-ray properties 
of elliptical galaxies 
by means of a chemical evolution code.
Detailed cooling and heating processes in the ISM
are taken into account using a mono-phase one-zone treatment, allowing us 
a reliable modelling of the galactic wind regime. 
For reproducing a giant elliptical we make use of the Pipino et al. (2005) case Ha1, which
is characterized by a luminous mass 
$M_{\rm lum}=10^{12}M_{\odot}$,
with a SF efficiency $\nu = 25\;\rm Gyr^{-1}$ and an infall timescale $\tau = 0.2$ Gyr,
in which a mild secular gas accretion from the surrounding IGM is allowed.
For a more typical spheroid ($L\sim L_*$) we refer to their model La1, 
which has a luminous mass $M_{\rm lum}=10^{11}M_{\odot}$, a 
SF efficiency $\nu = 15\;\rm Gyr^{-1}$ and an infall timescale $\tau = 0.3$ Gyr. 
We stress that, due to the uptodate nucleosynthesis prescriptions and to the inclusion
of dust, the models presented here are intended as an improved version of the original Pipino et al. (2005)
models Ha1 and La1.
To study the properties of a Lyman Break Galaxy, we use a model charactherized by a luminous mass 
$M_{\rm lum}=10^{10}M_{\odot}$, a star formation efficiency  $\nu = 5\;\rm Gyr^{-1}$ and an infall timescale $\tau = 0.5$ Gyr, 
referred to as LBGa1 model. In our scheme, large galaxies form the bulk of their stars in a shorter timescale than small galaxies, 
according to the ``inverse wind'' scenario. 
As shown by Matteucci (1994), this scenario allows us to 
reproduce the correlation between the [Mg/Fe] ratio and the galactic mass 
observed in local ellipticals.

In panel (b), we show the predicted SF history for two models of elliptical galaxies (La1 and Ha1), 
characterized by very high SFR values (from $\sim$ 200 to $\sim 3000$  $M_{\odot}/yr$) and by a 
starburst lasting $\sim 0.4-0.9$ Gyr. After this strong event, the SF stops and a galactic wind sets in for several Gyrs, 
until when a diffuse and hot gaseous halo, which surrounds the galaxy, is formed (for a more detailed description of this model, see 
Pipino et al. 2005).\\
Finally, in panel (c) of Fig.~\ref{SFR}, we show two different models tested for an irregular galaxy. The solid line represents 
a dwarf irregular 
with continuous SF (IC), whereas the dotted line is a starburst irregular.  
For the IC model, we assume a continuous star formation with SF efficiency of $\nu_{IC}=0.05$ Gyr$^{-1}$. 
On the other hand, the starburst model is similar to the model B used by Lanfranchi \& Matteucci (2003). The SFR of this model consists of 3 
bursts with an efficiency  $\nu_{burst}=0.5 Gyr^{-1}$. The times of occurrence of the bursts are 1 Gyr, 10 Gyr and 13 Gyr. The durations for each burst 
are 0.02 Gyr, 0.02 Gyr and 0.2 Gyr, respectively (for further details, see Lanfranchi \& Matteucci 2003).   
The IC and burst models have been used to reproduce the abundance patterns observed in Damped Lyman Alpha (DLA) systems 
(Calura et al. 2003, Dessauges-Zavadsky 
et al. 2004) and allow us to reproduce the present-day features of local dwarf irregulars and blue compact galaxies 
(Recchi et al. 2002, Lanfranchi \& Matteucci 2003), respectively. 
This latter model should be regarded as representative of a blue compact dwarf galaxy. Our assumption is based on the results 
by Lanfranchi \& Matteucci (2003), who have shown that the N/O, C/O, Si/O and O/Fe ratios observed in BCGs can be explained 
by a model with two to seven short bursts of SF with efficiencies in the range $\nu= 0.1 - 0.9 Gyr^{-1}$\\

\renewcommand{\baselinestretch}{1.0}
\begin{table*}
\centering
\begin{tabular}{lccccc}
\\[-2.0ex] 
\hline
\\[-2.5ex]
\multicolumn{1}{l}{}&\multicolumn{2}{c}{}&\multicolumn{1}{c}{Parameters}&\multicolumn{2}{c}{}\\
\hline
\multicolumn{1}{c}{Model}&\multicolumn{1}{c}{$\nu$}&\multicolumn{1}{c}{$\tau_{inf}$}&\multicolumn{1}{c}{$\sigma_{tot}$}&\multicolumn{1}{c}{$k$}&\multicolumn{1}{c}{IMF}\\
\multicolumn{1}{c}{}&\multicolumn{1}{c}{(Gyr$^{-1}$)}&\multicolumn{1}{c}{(Gyr)}&\multicolumn{1}{c}{($M_{\odot}/pc^{2}$)}&\multicolumn{1}{c}{}&\multicolumn{1}{c}{}\\
\hline
\hline
\\[-1.0ex]
Milky Way, S.N.      &    1     &    7     &      54        & 1.5   &  Scalo        \\
\hline
Milky Way, 16 Kpc    &    1     &    20     &      5        &  1.5  &  Scalo        \\
\hline
\hline
\multicolumn{1}{c}{Model}&\multicolumn{1}{c}{$\nu$}&\multicolumn{1}{c}{$\tau_{inf}$}&\multicolumn{1}{c}{$M_{lum}$}&\multicolumn{1}{c}{$k$}&\multicolumn{1}{c}{IMF}\\
\multicolumn{1}{c}{}&\multicolumn{1}{c}{(Gyr$^{-1}$)}&\multicolumn{1}{c}{(Gyr)}&\multicolumn{1}{c}{($M_{\odot}$)}&\multicolumn{1}{c}{}&\multicolumn{1}{c}{}\\
\hline
\hline
Elliptical, La1     &    15     &    0.3   &     10$^{11}$  & 1  &  Salpeter     \\
\hline
Elliptical, Ha1     &    25    &    0.2   &    10$^{12}$    & 1  &  Salpeter     \\
\hline
\hline
Irregular, IC       &   0.05   &    10    &     10$^{9}$    & 1  &  Salpeter     \\
\hline
Irregular, burst    &   0.5    &    10    &     10$^{9}$    & 1  & Salpeter     \\
\hline
\hline
\end{tabular}
\label{models}
\caption{}
Adopted parameters for the galactic models used in this work. 
The various models  are listed in column 1. In  
columns 2, 3, 4, 5 and 6  we present the adopted parameters, i.e.  the SF efficiency $\nu$, the infall timescale $\tau_{inf}$, 
the total surface mass density $\sigma_{tot}$ (in the case of the Milky Way) or the baryonic mass  $M_{lum}$ 
(in the case of the elliptical/irregular model), the Schmidt law exponent $k$ and 
the IMF, respectively. 
\end{table*}

\subsection{Chemical evolution of the Dust}
The chemical evolution of an element $i$ in the dust is computed by using 
the formalism developed by D98. 
Let $X_{dust, i} (t)$ be the abundance by mass of the element \emph{i} in the dust and 
$G(t)$ the ISM fraction at the time $t$, the quantity 

\begin{equation}
G_{dust,i}(t)= X_{dust,i} \cdot G(t)
\end{equation}

represents  the normalized mass density
of the element \emph{i} at the time \emph{t} in the dust. The time evolution 
of $G_{dust,i}(t)$ is calculated by means of the following equation:

\begin{eqnarray}
 & & {d G_{dust,i} (t) \over d t}  =  -\psi(t)X_{dust,i}(t)\nonumber\\
& & + \int_{M_{L}}^{M_{B_m}}\psi(t-\tau_m) \delta^{SW}_{i}
Q_{mi}(t-\tau_m)\phi(m)dm \nonumber\\ 
& & + A\int_{M_{B_m}}^{M_{B_M}}
\phi(m)\nonumber \\
& & \cdot[\int_{\mu_{min}}
^{0.5}f(\mu)\psi(t-\tau_{m2}) \delta^{Ia}_{i}
Q_{mi}(t-\tau_{m2})d\mu]dm\nonumber \\ 
& & + (1-A)\int_{M_{B_m}}^
{8 M_{\odot}}\psi(t-\tau_{m})  \delta^{SW}_{i} Q_{mi}(t-\tau_m)\phi(m)dm\nonumber \\
& & + (1-A)\int_{8 M_{\odot}}^
{M_{B_M}}\psi(t-\tau_{m})  \delta^{II}_{i} Q_{mi}(t-\tau_m)\phi(m)dm\nonumber \\
& & + \int_{M_{B_M}}^{M_U}\psi(t-\tau_m)  \delta^{II}_{i} Q_{mi}(t-\tau_m) 
\phi(m)dm \nonumber\\ 
& & - \frac{G_{dust,i}}{\tau_{destr}} + \frac{G_{dust,i}}{\tau_{accr}} -({d G_{dust,i} (t) \over d t})_{\rm out}  	
\label{eq_dust}
\end{eqnarray}

The main differences between eq.s ~\ref{eq_chem} and ~\ref{eq_dust} concern all the integrals in the 
right sides of both equations. These integrals,  
in the case of eq.~\ref{eq_dust} contain the quantities  $\delta^{SW}_{i}$, $\delta^{Ia}_{i}$ and  $\delta^{II}_{i}$, i.e. the condensation 
efficiencies of the element $i$ in stellar winds, type Ia and type II SNe (see section 2.2.1). 
These quantities represent the 
fractions of the element $i$ which is condensed into dust and restored into the ISM by low and intermediate 
mass stars, type Ia and type II SNe, respectively. 
The seventh and eighth terms of eq.~\ref{eq_dust} represent  the dust destruction and accretion rates, respectively. 
These terms depend on the quantities  $\tau_{destr}$ and $\tau_{accr}$, which represent the typical 
timescales for destruction and accretion, 
respectively. These two quantities are discussed in detail in sections 2.2.2 and 2.2.3, respectively. 
Finally, the last term of eq.~\ref{eq_dust}  accounts for possible ejection of dust 
into the inter galactic medium (IGM) by means of galactic winds. 
This term is absent in the equation for the S.N. model, 
but is taken into account in the elliptical and irregular models.

\subsubsection{Dust condensation efficiencies} 
We assume the dust condensation efficiencies suggested by D98. 
The choice of the following dust condensation efficiencies is motivated by various 
lines of evidence, as discussed in detail in D98. 
In this paper, 
we focus on the main refractory elements which are depleted into dust 
in the cold phase. These elements are C, O, Mg, Si, S, Ca, Fe. 
As an assumption, we consider that only these elements can be incorporated into dust grains. 
We assume that dust grains can be of two types: carbon (C) dust and silicate (Si) dust. 
We are aware that dust composition can be more complicated. For instance, Mathis (1996) presented 
a model including also the possibility of composite grains containing both carbon, silicates and oxides. 
Li \& Greenberg (1997) presented a trimodal dust model with large silicate core-organic refractory mantle dust particles, 
small carbonaceous particles and Polycyclic aromatic hydrocarbon particles, able to reproduce the interstellar 
extinction and polarization observational constraints. 
Furthermore, different dust structure would also have an impact on dust destruction and accretion (see Greenberg \& Li 1999, Jones et al. 1996). 
However, the inclusion of more complicated dust types is beyond the aims of this paper, since it  would increase the number of free parameters 
involved in our study. 
According to the formalism developed in D98, the contributors to 
the dust production are (a) low and 
intermediate mass stars, (b) type Ia SNe and  (c) type II SNe. We neglect the contributions from Wolf-Rayet stars and novae (D98), 
both of which are 
believed to represent unimportant sources of dust production.

\emph {a. Low and intermediate mass stars} - In these stars, dust is produced during the Asymptotic Giant Branch (AGB) phase 
(Ferrarotti \& Gail 2006 and references therein).  We assume that 
dust formation depends mainly on the composition of the stellar envelopes. 
If $X_{O}$ and $X_{C}$ represent the O and C mass fractions in the stellar envelopes, respectively, we assume that 
stars with $X_{O}/X_{C}$ $>$ 1 are producers of silicate dust, i.e. dust particles composed by O, Mg, Si, S, Ca, Fe. On the other hand, C rich stars, 
characterized by $X_{O}/X_{C}$ $<$ 1, are producers of carbonaceous solids, i.e. carbon dust (Draine 1990). 
Being $M_{i, ej}(m)$ and $M_{i, dust}(m)$ the total ejected mass and the dust mass formed by the stars as functions of the initial mass $m$ 
for the element $i$, 
respectively, we assume that for stars with $X_{O}/X_{C}$ $<$ 1 
\\
\\
\begin{math}
M_{dust, C}(m) =  \delta^{SW}_{C} \cdot [M_{C, ej}(m)-0.75 M_{O, ej}(m)]\\
\end{math}
\\
with $\delta^{SW}_{C}=1$ and 
\\
\\
\begin{math}
M_{dust, i}(m) =0, \\
\end{math}
\\
for all the other elements. For stars with $X_{O}/X_{C}$ $>$ 1 in the envelope, we assume 
\\
\\
\begin{math}
M_{dust, C}(m) =0 \\
\end{math}
\\
\\
\begin{math}
M_{dust, i}(m) = \delta^{SW}_{i} M_{i, ej}(m) \\
\end{math}
\\
\\
with $\delta^{SW}_{i}=1$ for Mg, Si, S, Ca, Fe and \\
\\
\\
\begin{math}
M_{dust, O}(m)=16 \sum_{i} \delta^{SW}_{i} M_{ej, i}(m)/\mu_{i}   \\
\end{math}
\\
\\
with $\mu_{i}$ being the mass of the $i$ element in atomic mass units.

\emph {b. Type Ia Supernovae} - for these systems, we assume 
\\
\\
\begin{math}
M_{dust, C}(m) = \delta^{Ia}_{C}[M_{ej, C}(m)]\\
\end{math}
\\
\\
with  $\delta^{Ia}_{C}=0.5$; 
\\
\\
\begin{math}
M_{dust, i}(m) = \delta^{Ia}_{i}M_{ej, i}(m)\\
\end{math}
\\
\\
with $\delta^{Ia}_{i}=0.8 $ for Mg, Si, S, Ca, Fe; \\
\\
\\
\begin{math}
M_{dust, O}(m)=16 \sum_{i} \delta^{Ia}_{i} M_{ej, i}(m)/\mu_{i}  \\
\end{math}

\emph {b. Type II Supernovae} - For type II SNe, we adopt the same prescriptions as for type Ia SNe.
In a recent paper, Zhukovska et al. (2007)  by means of a chemical evolution model 
for the solar neighbourhood 
study the available observations of presolar dust grains in meteorites. 
Their analysis points towards condensation efficiencies lower than the ones suggested by D98, in particular 
concerning the silicates. In this paper, we test various assumptions concerning these parameters 
and study the effects also on systems other than the solar neighbourhood.

\subsubsection{Dust Destruction}

Dust destruction is primarily due to the propagation 
of SN shock waves in the warm/ionized interstellar medium (McKee 1989, Jones et al. 1994). 
Evidences for dust destruction come from the observations of high velocity clouds, where  
an anti-correlation between the depletion levels and the cloud velocities 
has been found (Shull 1978, McKee 1989). This was interpreted as an 
evidence for grain destruction in SN shocks (D98). 
Following the suggestions by McKee (1989) and D98, for a given element $i$ 
the destruction timescale $\tau_{destr}$ can be expressed as: 
\begin{equation}
\tau_{destr, i}=(\epsilon M_{SNR})^{-1} \cdot \frac{\sigma_{gas}}{R_{SN}}
\end{equation} 

Hence the destruction timescale is independent from the dust mass.  
$M_{SNR}$ is the mass of the interstellar gas swept up by the SN remnant. 
For this quantity, McKee (1989) suggests 
a typical value of $ M_{SNR} \sim 6800 M_{\odot}$, which is in agreement 
with the results from numerical studies of SN evolution (Thornton et al. 1998). 
As suggested by McKee (1989), typical 
values for the destruction efficiency $\epsilon$  in a three-phase medium 
as the present-day local ISM are around 0.2, hence we assume:  
\begin{equation}
\epsilon M_{SNR} = 0.2 \times 6800 M_{\odot} = 1360 M_{\odot}
\, .
\end{equation}
$R_{SN}$ is the total SNe rate, 
including the contributions by both type Ia and type II SNe. 
No significant modifications are required in a single-phase gas modelling (as we did in this paper,
 McKee1989). 
 We note, however, that McKee (1989) estimates $\epsilon M_{SNR} \sim 70 M_{\odot}$ for a hot and rarefied medium as the gaseous halos
surrounding ellipticals, therefore we run two models in which $\epsilon M_{SNR} = 1360 M_{\odot}$ until the galactic wind
and then we have an instantaneous transition to $\epsilon M_{SNR} \sim 70 M_{\odot}$. The pre-wind
dust evolution is obviously unaffected, whereas it leads to substantial changes in the late
stage of the galactic evolution. According to the galaxy mass, we call them \emph{La1+MK} and \emph{Ha1+MK}, respectively, and we will show that
they might produce interesting results, even though they use an over-simplified
treatment of the dust destruction in the hot gas. 
Finally, in order to render the cases \emph{La1} and \emph{Ha1} more realistic, 
we further modified the destruction treatment by implementing a thermal sputtering term,
which is thought to be the dominant source of dust destruction in hot plasmas.
In particular, following Itoh (1989),
we assume that, in a $\sim$1 keV plasma, nearly 90\% of the dust grains will evaporate
by thermal sputtering in $\tau_{destr_{sp}, i} \sim 10^5 /n_e (\rm yr\; cm^{-3})$, 
where the electron density $n_e$ has been self-consistently evaluated at each timestep.
This translates into a new destruction term, namely: 
\begin{equation}
\frac{G_{dust,i}}{\tau_{destr}}= G_{dust,i} (70 M_{\odot}) \frac{R_{SN}}{\sigma_{gas}} + \frac{G_{dust,i}}{\tau_{destr_sp,i}}
\end{equation} 
According to the galaxy mass, we call the models featuring this particular term as \emph{La1+Itoh} and \emph{Ha1+Itoh}, respectively.
For elliptical galaxies, these two models will be regarded as the \emph{fiducial} ones. 
According to Itoh (1989), this term dominates the Fe grains evolution at late times. 

In section 3.1.2, we test the parameter $\epsilon$ and we tune it 
in order to reproduce the depletion pattern observed in 
the S.N. 
Further refinements are presented in section 3.2. 

\subsubsection{Dust Accretion}

Dust accretion occurs in dense molecular clouds, 
where volatile elements can condensate onto pre-existing grain cores, originating a volatile 
part called mantle (D98, Inoue 2003).  Direct evidences for dust accretion come from the observed  
large variations of the depletion levels as a function of the density (Savage \& Sembach 1996) and from the observed 
 infra red emission of cold molecular clouds (Flagey et al. 2006), which is characterized by the absence 
of small grain emission. 
These features can be accounted for by the coagulation of small grains on and into larger particles. 
Indirect evidence for dust accretion 
comes from the estimation of the grain lifetimes, 
which would be very small if no process could allow the grains to recondense and grow (McKee 1989, Draine \& Salpeter 1979). 
For a given element $i$,  
the accretion timescale $\tau_{accr}$ can be expressed as: 

\begin{equation}
\tau_{accr}=\tau_{0,i}/(1 - f_i) 
\label{accr_t}
\end{equation} 

where 
\begin{equation}
f_i=\frac{G_{dust,i}}{G_{i}}
\end{equation}

According to eq.~\ref{accr_t}, the accretion timescale is an increasing function of the dust mass. 
For the timescale $\tau_{0,i}$, typical values span from $\sim 5 \times 10^{7}$ yr, of the order of 
the lifetime of a typical molecular cloud (D98), up to $\sim 2 \times 10^{8}$ yr  (D98). 
In this paper, we assume that the timescale $\tau_{0,i}$ is constant for all elements, 
with a value of $5 \times 10^{7}$ yr.\\
In elliptical galaxies, we assume that dust accretion occurs only during the starburst epoch, when large 
amounts of cold gas and 
molecular H are available. After the onset of the wind and the end of the starburst, 
we assume that no molecular gas is present, hence no more accretion can occur.\\
The observed molecular H content in dwarf irregular galaxies is very small, with molecular-to-atomic gas fractions 
of $\sim 10 \%$ or lower (Lisenfeld \& Ferrara 1998, Clayton et al. 1996). 
Motivated by these observational results, we assume that no accretion can occur in irregular galaxies. 

\section{Results}
\subsection{Dust evolution in the Solar Neighbourhood}

\subsubsection{Dust Production Rates}
\begin{figure*}
\centering
\vspace{0.001cm}
\includegraphics[height=18pc,width=18pc]{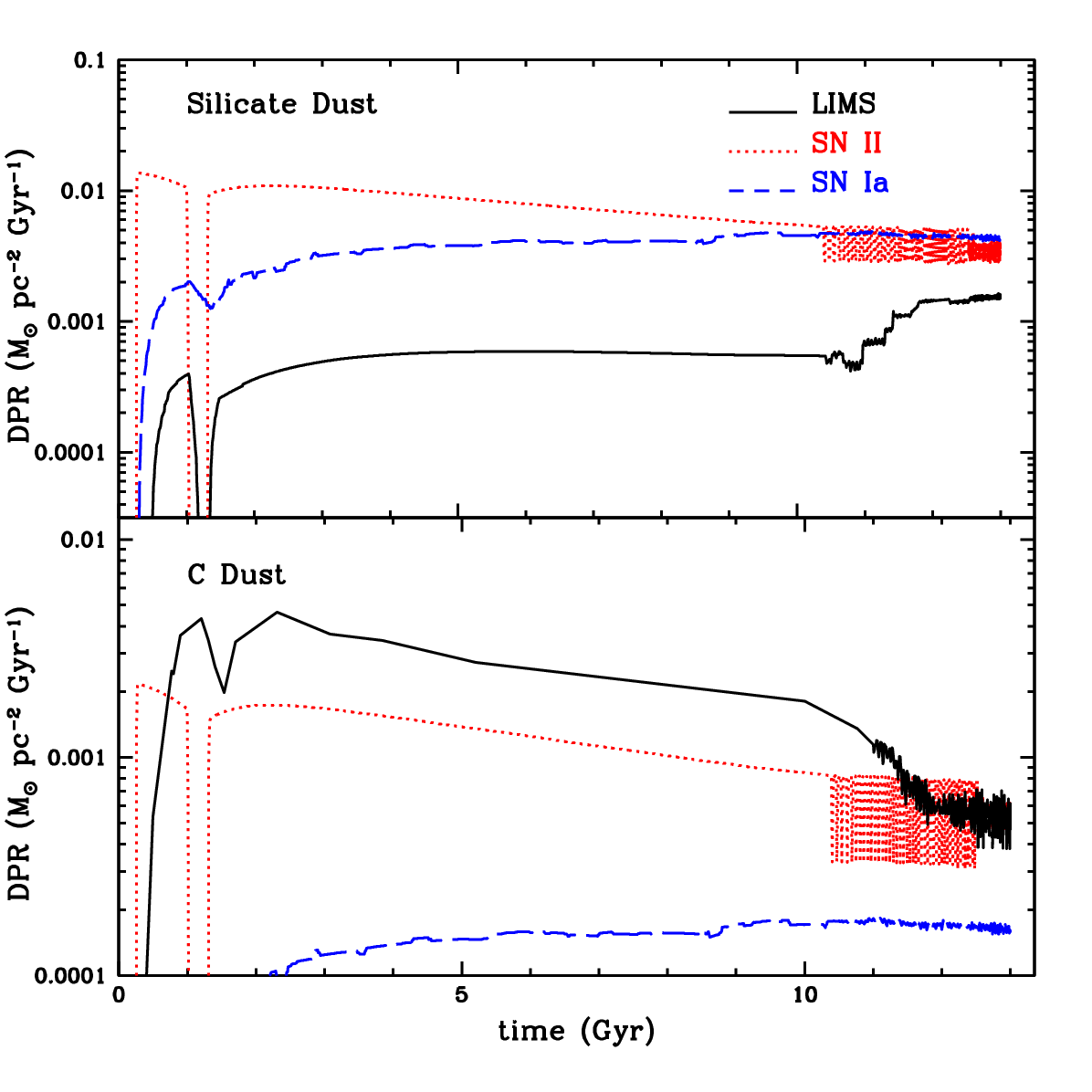}
\caption[]{Predicted dust production rates from various sources 
for a chemical evolution model of the S.N.  
In the lower (upper) panel, we show the results for the carbon (silicate) dust. 
Solid lines: contribution by low and intermediate mass stars (LIMS). Dotted lines: contribution 
by type II SNe. Dashed lines: contribution by type Ia SNe. }
\label{prates}
\end{figure*}

In Figure~\ref{prates}, we show the predicted evolution of the carbon (lower panel) and silicate (upper panel)  
dust production rates calculated by means of the chemical evolution model for the S.N. 
In particular, we show the contributions to the rates 
by various sources: the low and intermediate mass stars (LIMS, solid lines), the type II SNe (dotted lines) and the 
type Ia SNe (dashed lines). Concerning the C dust, its production is dominated by LIMS throughout most of the cosmic time. 
At the present day ($T_{0}\sim 13$ Gyr) a significant contribution is coming also by type II SNe, with type Ia SNe 
playing  a negligible role. The frequent discontinuities in the lines are due to the effect of the star 
formation threshold.\\
The production of silicate dust is dominated by type II SNe throughout most of the time. At late times, the contributions by type Ia SNe 
and  type II SNe are comparable. On the other hand, LIMS are negligible contributors to the Si dust production rate.

\subsubsection{The dust fractions}

An important test for the robustness of our formalism is represented by the study of the metal fractions in dust. 
These quantities are the ratios between the amount of a given element locked into dust 
and its total abundance and can be expressed as: 

\begin{equation}
f_{i} = \frac{X_{dust,i}}{X_{i}}
\end{equation}

\begin{figure*}
\centering
\vspace{0.001cm}
\includegraphics[height=18pc,width=18pc]{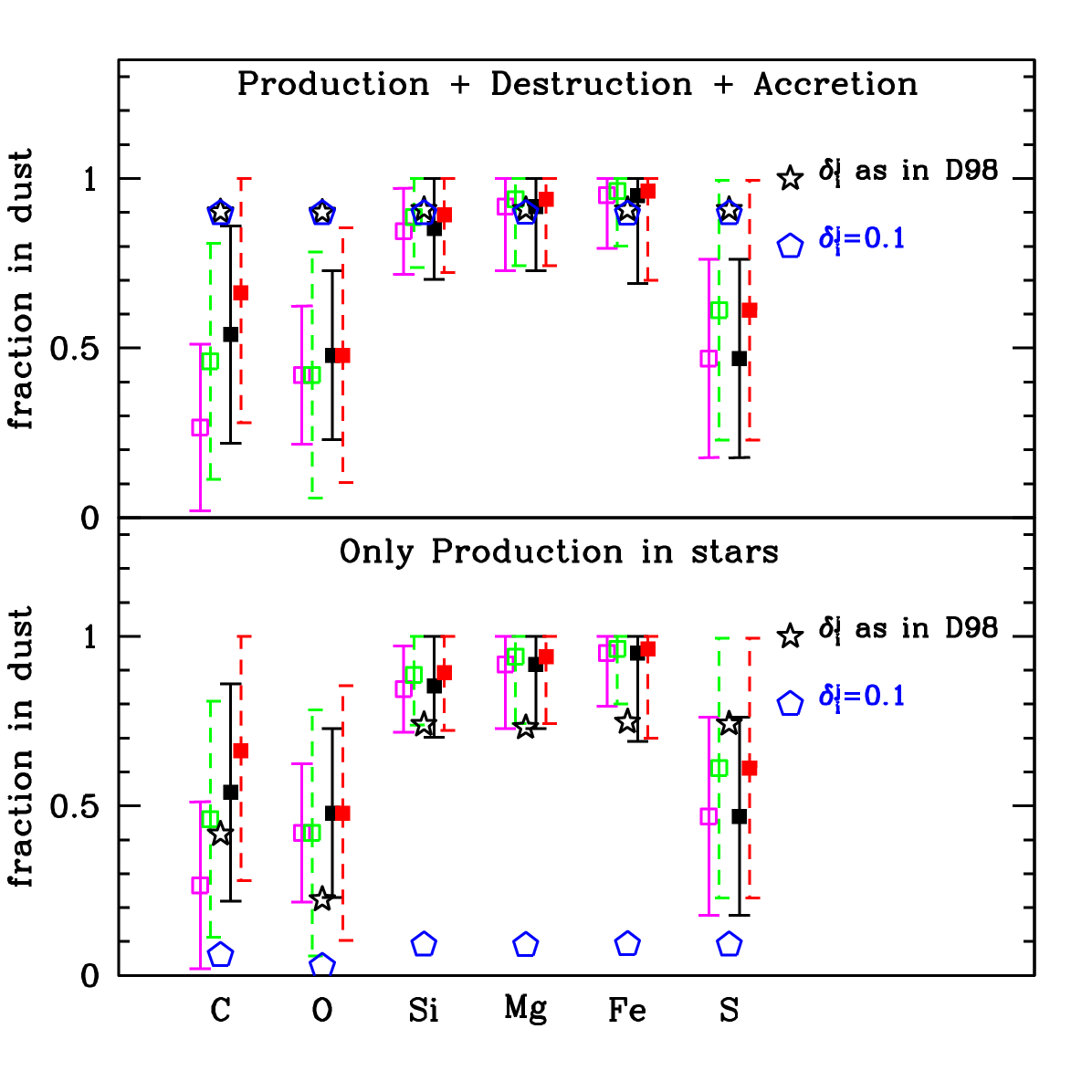}
\caption[]{Fractions in dust for various elements. 
Open stars: predicted present-day fractions calculated by adopting for the dust condensation efficiencies 
the prescriptions suggested by D98.
Open pentagons: predicted present-day fractions calculated by assuming for the dust condensation efficiencies 
a constant value of 0.1. The solid and open squares are the fractions observed by Kimura et al (2003) in the Local Interstellar 
Cloud using the set of cosmic abundances specified in their table 2 and 3, respectively. 
The squares with solid and dashed error bars have been calculated assuming a H ionization fraction of $\chi_{H}=0.25$ and 
$\chi_{H}=0.45$, respectively. In the lower panel, the predicted values are calculated considering only dust production in stars. 
In the upper panel, the predicted values are calculated considering dust production in stars, dust destruction by SNe and dust 
accretion in the ISM.}
\label{fract_delta}
\end{figure*}

Of particular interest is the comparison between the predicted dust fractions for various elements and 
the observed ones. This analysis is very useful to test the various parameters involved in our study, 
in particular the dust condensation efficiencies $\delta^{SW}_{i}, \delta^{Ia}_{i}$, and $\delta^{II}_{i}$, 
the dust destruction efficiency 
$\epsilon$ and the dust accretion timescale  $\tau_{0,i}$, as well as the sensitivity of the results on 
the assumptions about these parameters. 
In Fig.~\ref{fract_delta} we show the predicted present-day fractions in dust for the 
elements studied in this work, compared to the values observed by Kimura, Mann \& Jesseberg (2003) in the Local Interstellar Cloud. 
It is worth to note that the Local Interstellar Cloud consists of a warm medium, characterized by a temperature of $\sim 6000 K$, 
whereas our models provide the depletion of the cold gas, with temperatures lower than $\sim 100 K$. 
In this work, the dust fractions observed by Kimura et al. (2003) are taken as reference. 
From the logarithmic depletions $\delta$ plotted in Figure 2 of Kimura et al. (2003), 
it is possible to derive the dust fractions of the cold gas through the formula 
\begin{equation} 
f=1-10^{\delta}
\end{equation} 
For C, O and S, the logarithmic depletions of the cold medium are very similar to the ones of the Local Intersetllar Cloud. 
For Si, Mg and Fe, the dust fractions of the cold medium are higher than the ones of the warm medium and are within the error bars 
plotted in Figure~\ref{fract_delta}. 
In the two panels of Fig.~\ref{fract_delta}, the predictions are obtained by testing two different assumptions concerning the dust condensation 
efficiencies. 
In the lower panel of Fig.~\ref{fract_delta}, we present the predicted values calculated considering only dust production in stars. 
The open stars are the fractions calculated by adopting for the dust condensation efficiencies 
the prescriptions suggested by D98. The open pentagons are the fractions calculated by assuming for the dust condensation efficiencies 
a constant value of 0.1, regardless of the element and of the source of production. 
Following the prescriptions suggested by D98, a higher amount of metals is locked into dust than with the other 
choice. The only source of dust destruction present in this case, i.e. astration, does not seem to play a dominant role in determining 
the dust depletion pattern, once destruction by SNe and accretion in the ISM are neglected. 
We outline that neglecting 
these two processes is unrealistic since, as stressed in sections 2.2.2 and 2.2.3, there are robust evidences that 
destruction and accretion play a non negligible role in the evolution of 
dust in the S.N. (see McKee 1989, Draine 1990, Tielens 1998).  
For purpose of comparison, we show also the observational values derived by Kimura et al. (2003) 
(open squares with solid and dashed error bars). 
The chemical composition of dust grains cannot be directly observed. 
The only way to estimate the abundance of an element locked up into dust is by means of its 
gas phase abundance, which is directly observable, and by means of an appropriate value representing the 
cosmic 
abundance. Observationally, for a given element $i$, 
the abundance in dust is derived  by means of the subtraction between the cosmic abundance 
$(X_{i,C})_{obs}$ and the 
observed gas abundances $(X_{gas,i})_{obs}$: 
\begin{equation}
(X_{i,dust})_{obs} = (X_{i,C})_{obs} - (X_{gas,i})_{obs}
\end{equation}

The observational determination of the dust fractions depends on the assumptions on the 
total cosmic abundances $(X_{i,C})_{obs}$ and on the H ionization fractions $\chi_{H}=0.25$. 
In the two panels of Fig.~\ref{fract_delta}, we consider four different sets of observed dust fractions 
(for details see caption of Fig.~\ref{fract_delta}), calculated with various assumptions of both $(X_{i,C})_{obs}$ 
and $\chi_{H}=0.25$. \\
In the upper panel of  Fig.~\ref{fract_delta}, the predicted values 
have been calculated by taking into account dust production in stars, 
dust destruction and dust accretion in the ISM. 
The calculation of dust destruction and accretion depends on the choice of two parameters, 
i.e. the destruction efficiency $\epsilon$ and the accretion timescale $\tau_{0,i}$, respectively. 
In this case, the predictions have been calculated by assuming  $\epsilon \sim 0.2$ and $\tau_{0,i}=5 \times 10^{7}$ yr for all elements. 
For this choice, the dust fractions are nearly independent from 
the choice of the dust condensation efficiencies.
This confirms the results by Dwek (1998), who showed that, at the present time, the accretion rate 
balances the destruction rate, and that the balance determines the depletion. 
Similar conclusions are drawn from the analysis of Zhukovska et al. (2007). 
By means of a different approach, 
Tielens (1998) attempted to determine the accretion and destruction rates from the observed depletions. 
The implication of these results is that, to derive constraints on the condensation efficiencies, one has 
to study the depletion pattern in systems where either destruction or accretion are absent.
In section 3.3.4, we will show that systems suited to this study are the dwarf irregular galaxies, where 
dust accretion is likely to play a negligible role. \\
From the upper panel of Fig.~\ref{fract_delta}, we note also that for some elements, the 
assumption of a constant destruction efficiency provides a very poor fit to the observed dust fractions. 
These elements are C, O and S, for which 
the predictions indicate present-time dust fractions of $\sim 0.9$, higher than the values observed 
in the Local Interstellar Cloud. 

\begin{figure*}
\centering
\vspace{0.001cm}
\includegraphics[height=18pc,width=18pc]{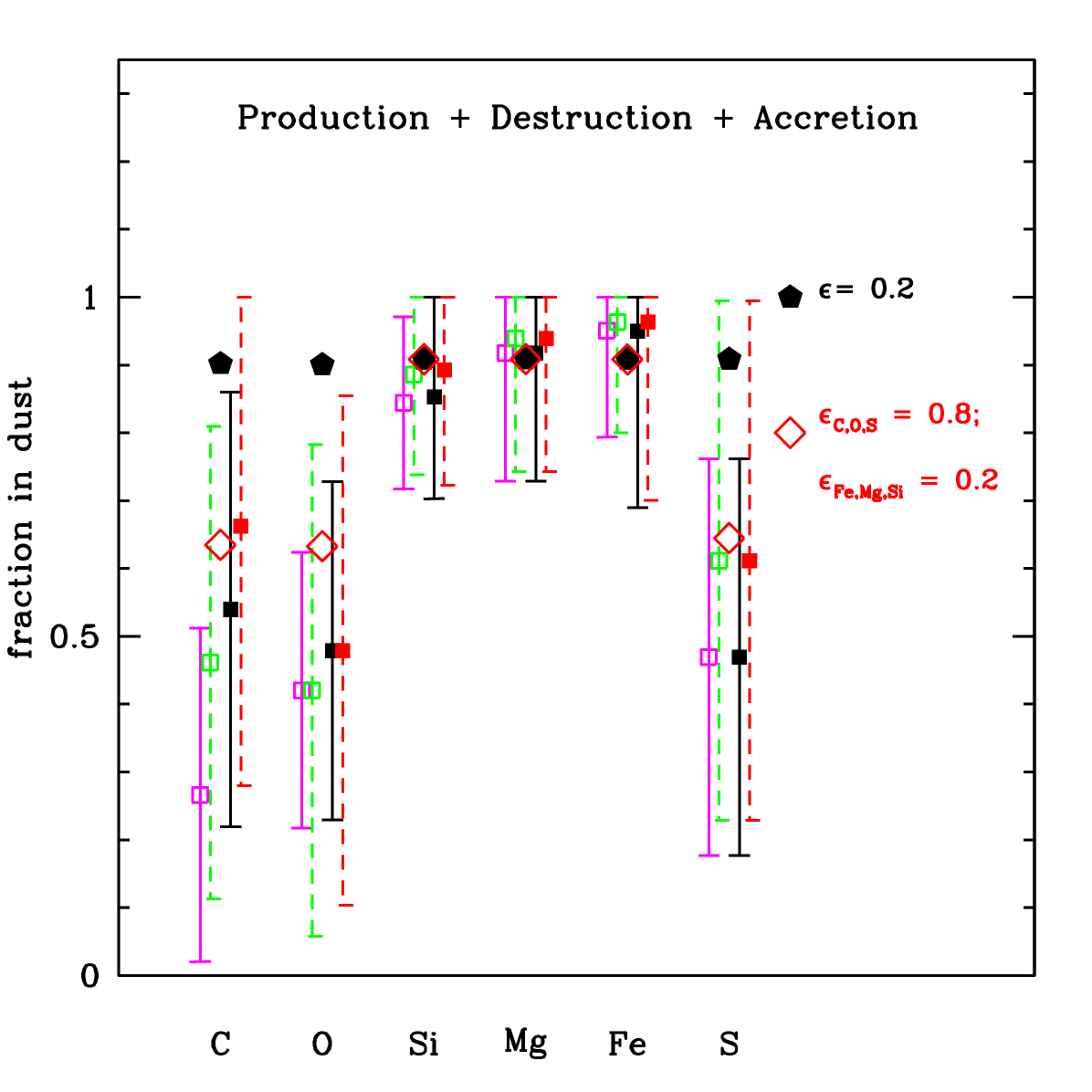}
\caption[]{Fractions in dust for various elements. 
The solid pentagons are the predicted present-day fractions calculated by assuming a 
dust destruction efficiency of $\epsilon=0.2$ for all elements.
The open diamonds are the present-day fractions calculated assuming a 
dust destruction efficiency of  $\epsilon_{C,O,S}=0.8$ for C, O and S and  whereas for 
$\epsilon_{Fe,Mg,Si}=0.2$ for Fe, Si and Mg. 
The squares with solid and dashed error bars are the same as in Figure~\ref{fract_delta}.}
\label{fract_epsi}
\end{figure*}

\begin{figure*}
\centering
\vspace{0.001cm}
\includegraphics[height=18pc,width=18pc]{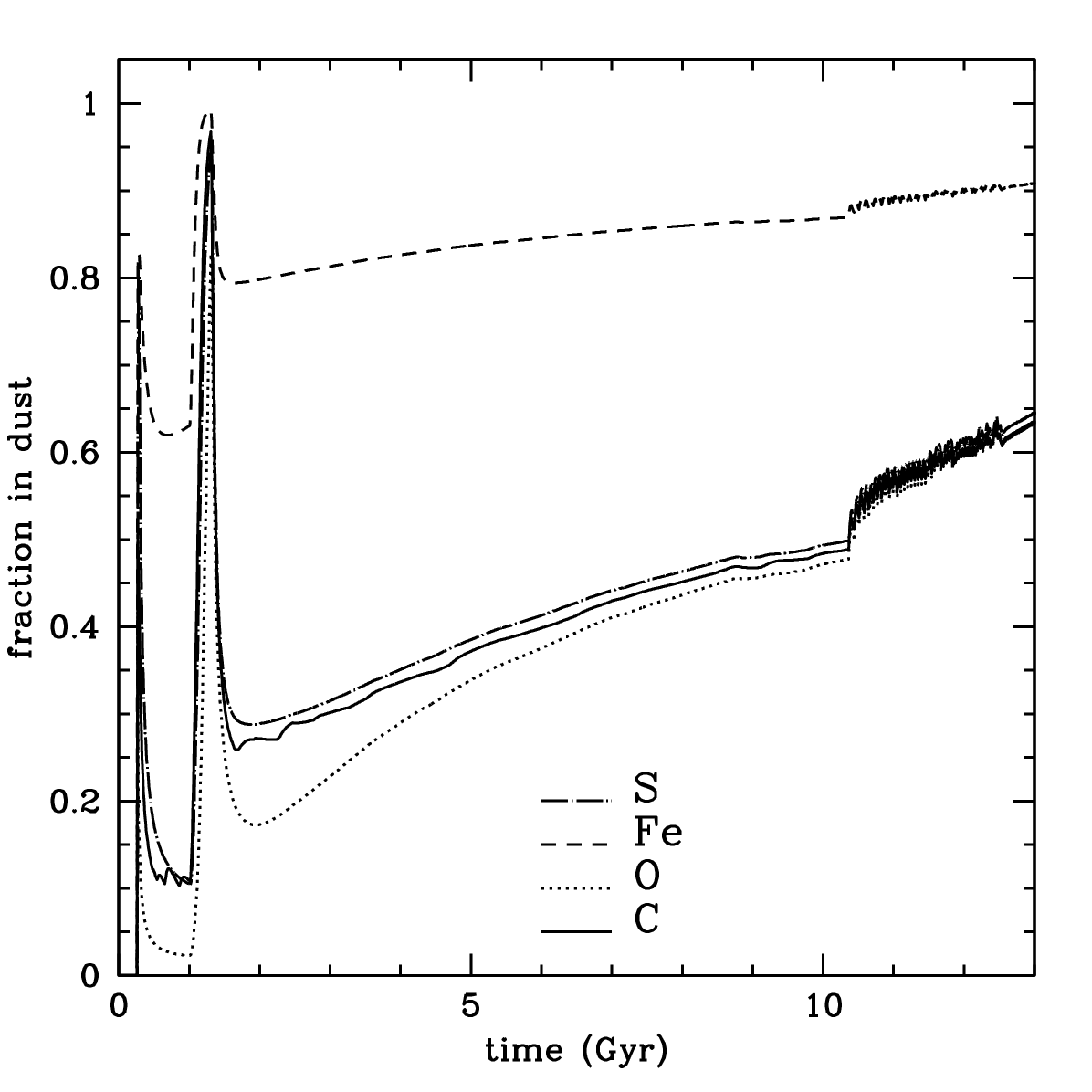}
\caption[]{Time evolution of the fractions in dust for various elements: C (solid line), O (dotted line), 
Fe (dashed line) and S (dot-dashed line). The evolution of the dust fractions for Mg and Si is 
identical to the Fe one.}
\label{fract}
\end{figure*}

The observed dust fractions can be reproduced in two ways, by assuming that either the dust 
accretion timescale or the 
the destruction efficiency depends on the physical properties of the chemical element. 
The dust accretion timescale is related 
to the micro-physics of the 
molecular clouds, which can not be modelled properly with our instruments. On the other hand, the destruction efficiency is 
connected to the heating of the ISM, mainly due to SNe and their explosion rate, which is accounted for in detail by our models.  
In this paper, we assume that the dust destruction efficiency depends on the properties of 
the chemical element. 
A physical justification is related to the condensation temperatures of the various elements. 
These quantities express 
the volatility of the elements in dust (Lodders 2003 and references therein). 
In general, elements with higher condensation 
temperatures  are more likely to aggregate into dust grains and are more resistant to destruction. 
Lodders (2003) has calculated the condensation temperatures for various elements for a solar-system composition gas. 
Her results indicate condensation temperatures of 78 K, 182 K and 704 K  for C, O and S, respectively. 
On the other hand, the condensation temperatures calculated for Fe, Si and Mg are 1357 K, 1529 and 1397, respectively, 
hence considerably higher than for C, O and S. 
Here, we assume that the  
destruction efficiencies vary as a function of the condensation temperatures of the elements. 
For C, O and S we assume a dust destruction efficiency of $\epsilon_{C,O,S}=0.8$, whereas for 
Fe, Si and Mg we assume  $\epsilon_{Fe,Mg,Si}=0.2$. 
As can be seen from Figure ~\ref{fract_epsi}, 
where we show the dust fractions calculated by assuming the above set of dust destruction efficiencies,   
with these values it is possible to reproduce the observed dust fractions with good accuracy. \\
In Figure ~\ref{fract}, we show the time evolution of the dust fractions for various elements (C, O, Fe and S). 
We have omitted the Mg and Si because for both elements, 
the evolution of the fraction in dust is identical to the one predicted for Fe. 
The curves for C, O and S show a similar behaviour after 
the minimum at 2 Gyr, with an increase up to the present time which is steeper than the one for Fe. 
This fact is due to the different destruction efficiencies adopted for the C, O and S group and  
for the Si, Fe and Mg group.

\subsubsection{The dust destruction and accretion rates} 
\begin{figure*}
\leftline{\includegraphics[height=18pc,width=18pc]{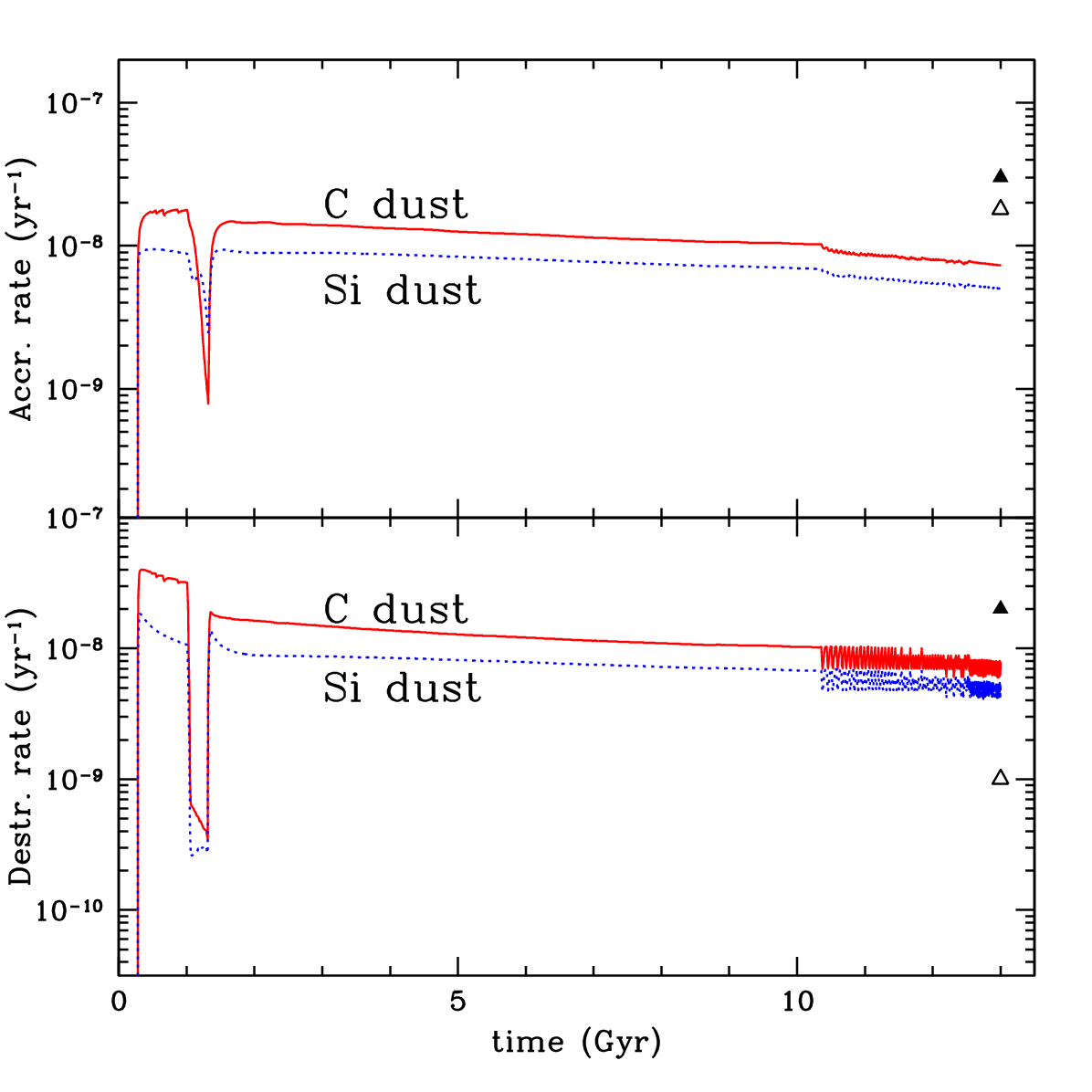}
\leftline{\includegraphics[height=18pc,width=18pc]{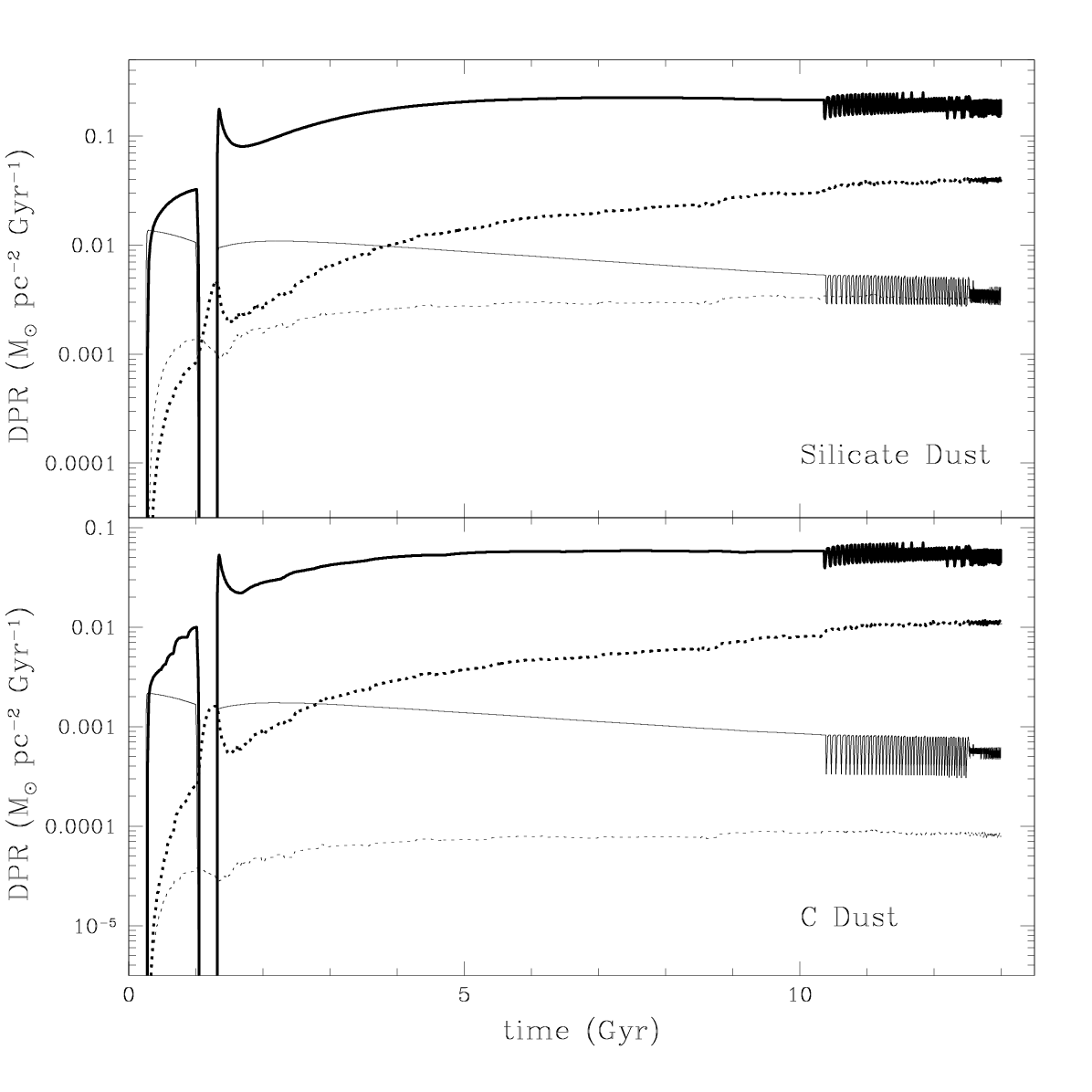}} } 
\caption[]{\emph{Left:} Evolution of the dust accretion (upper panel) and destruction (lower panel) rates in the S.N. 
The solid and dotted lines represent the calculated rates for carbon and silicate dust, respectively. 
The open triangles and solid triangles  are the present-day average values estimated by Draine (1990) and Tielens (1998), 
respectively, for a cold neutral medium. \emph{Right:} Evolution of the production and destruction rates by SNe  
for C dust (lower panel) and Si dust (upper panel). The thin dotted and solid lines are the 
predicted dust production rates by type Ia and type II SNe, respectively. 
The thick dotted and solid lines are the predicted dust destruction rates 
by type Ia and type II SNe, respectively. }
\label{rates_SN}
\end{figure*}

Once we have tuned the destruction efficiency  $\epsilon$ and chosen a realistic value of the accretion timescale $\tau_{0,i}$, 
we study the evolution of the accretion and destruction 
rates and we compare our values to the ones estimated by other authors. In the left panel of figure ~\ref{rates_SN}, we show the  
predicted evolution of the dust destruction and accretion rates for the S.N. 
The evolution of both rates is sensitive to the SF and gas accretion history of the model for the 
S.N. 
In the literature, no observational estimate can be found for the accretion or destruction rates in the S.N. 
These quantities have been calculated by various authors on the basis of theoretical investigations 
(McKee 1989, Draine 1990, Jones et al. 1994, Tielens 1998 ). 
Our model is one-phase and represents one single medium dominated by cold, neutral gas. 
The rates predicted by our models with the average estimates by Draine (1990) and Tielens 
(1998)  for a cold neutral medium. The values by Draine (1990) and Tielens (1998) are order of magnitude estimations, 
and are provided with no error bar. For the dust destruction rate, we predict present-day values of $\sim 5 \times 10^{-9}$yr$^{-1}$
and  $\sim 8 \times 10^{-9}$yr$^{-1}$ for silicate and carbon dust, respectively. Similar values are predicted for the accretion rates 
for both dust types. Our values are in good agreement with the order-of-magnitude estimations determined by Draine (1990) and 
Tielens (1998) with different approaches.

Several recent observational and theoretical results have addressed the role of SNe in dust production 
in the Milky Way (Dunne et al. (2003) and in high redshift galaxies 
(Hughes et al. 1998, Bertoldi et al. 2003, Morgan \& Edmunds 2003, Maiolino et al. 2004). 
Beside producing dust, SNe are the main sources of energy injection into the ISM and their hot cavities 
are known to destroy dust grains (D98). Our aim here is to compare the type Ia and II 
SN destruction and production rates and see how do these quantities evolve with cosmic time. 
In the right panels of Figure~\ref{rates_SN}, we study the balance between production and destruction by SNe 
for C dust (lower panel) and Si dust (upper panel). 
The thin dotted and solid lines are the 
predicted evolution of the dust production rates 
by type Ia and type II SNe, respectively.  
The thick dotted and solid lines are the 
predicted evolution of the dust destruction rates 
by type Ia and type II SNe, respectively.  
For type II SNe, 
dust destruction dominates over dust production 
throughout almost all the cosmic history. 
Only for a very short time ($\sim 0.3$ Gyr after the beginning of SF), 
the type II SN production rate was larger than the type II SN destruction rate. 
This is in agreement with the results by D98, who 
found that at earlier epochs, when the ISM metallicity was below $Z\sim 0.001$ 
(reached, according to our model, in $\sim 0.35$ Gyr) 
SNe inject more dust into the ISM than they can destroy. \\
On the other hand, for type Ia SNe 
during the first Gyr of galactic evolution, Si dust production has dominated over 
dust destruction, but with negligible consequences on the total Si dust mass. 
We conclude that, according to our results, throughout the whole history of the S.N., the presence of SNe, 
in particular of type II SNe, is fundamental for the process of cycling the dust back into the ISM.

\begin{figure*}
\centering
\vspace{0.001cm}
\includegraphics[height=18pc,width=18pc]{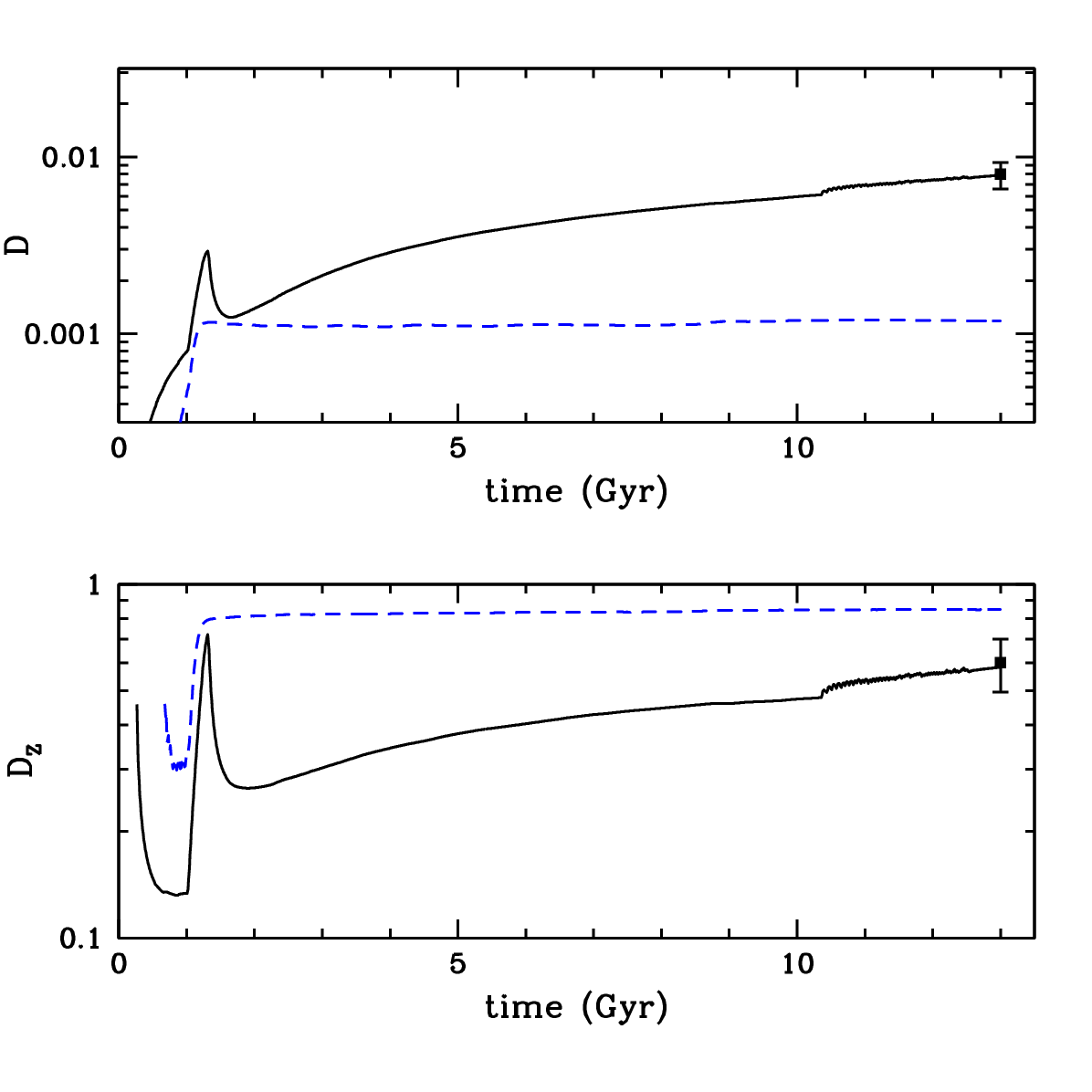}
\caption[]{Upper panel: predicted evolution of the dust to gas ratio in the S.N. (solid line) and 
at 16 Kpc from the Galactic centre (dashed line). 
The solid square with error 
bar is the observational value for the Galaxy taken from Issa et al. (1990). 
Lower panel: predicted evolution of the dust to metals ratio in the S.N. (solid line) and 
at 16 Kpc from the Galactic centre (dashed line). The open square with error 
bar is the observational value for the Galaxy assuming the dust to gas ratio estimated by Issa et al. (1990) and a solar metallicity 
of $Z_{\odot}=0.0133$ (Lodders 2003).  }
\label{dtg_dtm}
\end{figure*}

\subsubsection{Dust to gas and Dust to metals ratios and their evolution}

We define the dust to gas ratio $D$ as the ratio between the dust mass
and the total ISM mass surface densities,

\begin{equation}
D = \frac{\sigma_{dust}}{\sigma_{ISM}}
\end{equation}
while the dust to metal ratio  $D_{Z}$ is the ratio between the dust
mass and the metal mass surface density in the ISM,

\begin{equation}
D_{Z} = \frac{\sigma_{dust}}{\sigma_{ISM} \cdot Z} = \frac{D}{Z}
\end{equation}

Issa,
MacLaren and Wolfendale (1990), by studying the $D$ quantity in local spirals,
have found a correlation  between $D$ and the metallicity. The
implication of this fact is that the dust to metals ratio is nearly
constant for local spirals.  In the upper panel of Figure~\ref{dtg_dtm},
we show the predicted time evolution of D for two different zones of our Galaxy: 
the S.N. (solid line) and 
a circular ring 1 kpc wide located  at a distance  of 16 Kpc from the Galactic center (dashed line). 
The latter represents a model for the outermost regions of the spiral disc (see table 1). 
The comparison between these two models 
is interesting to understand the differences of the dust properties for two regions located in different 
zones of the Galactic disc. 
We plot also the Galactic value  (solid square with error
bars) $D_{\odot} = 0.008 $, taken from Issa et al. (1990), which is well reproduced by our
model. 

In the lower panel of Figure~\ref{dtg_dtm},
we plot the time evolution of the dust to metals ratio $D_{z}$ for the S.N. (solid line) 
and the spiral outskirts (dashed line) models. 
We compare  the predictions to an observational value derived for the Galaxy. 
By assuming for the S.N. a metallicity of $Z_{\odot} = 0.0133 $ (Lodders 2003),  
for our Galaxy we obtain $D_{Z,\odot}=0.6$. Interestingly, this value coincides 
with our predicted dust fraction for O (see Fig.~\ref{fract}), which is the element dominating the total metallicity.\\
Chemical
evolution studies have shown that the dust to metals ratio should not 
vary much during galaxy evolution (Inoue 2003, Edmunds 2001).  Our 
predictions indicate that in the initial phases of galaxy evolution,
in particular from the halo phase to the SF hiatus, 
$D_{z}$ experiences strong variations, evolving rapidly  from 0.15 to
0.8. After the peak at 1 Gyr, $D_{z}$ drops sharply
and its evolution in the disc phase is much slower,  i.e. of a factor
of $\sim 2$ from 2 Gyr up to the present time.  This is consistent
with the results by Inoue (2003).  The present-day value is nicely
reproduced by our models. The spiral outskirt model presents 
$D_{z}$ values  higher than the S. N. model. 
Dust accretion is likely to take place in the outermost regions of the Milky Way owing to the presence of molecular clouds. 
In fact, recent near Infrared observations have revealed the presence of molecular clouds in regions located at galactocentric 
distances out to $R \ge 13.5-20 $ kpc (Kobayashi \& Tokunaga 2000, Snell et al. 2002, Nakagawa et al. 2005). \\

 In table 2, we present a detailed comparison between 
the numeric values as predicted by means of our model for the S.N. with the predictions by D98, 
as well as with the available 
corresponding observational results. 
Concerning the predicted ISM properties, our results are very similar to the ones of D98 and well 
within the available observational values. The small differences are due mainly to different 
prescriptions for the SF law (see our equation 5 and eq. 15 of D98) and for the infall law, 
for which we assume the two infall expression by Chiappini et al. (1997), whereas D98 assumes a simple 
exponential law. \\
Differences in the C and Si dust production rates from LIMS and SNe (Ia+II) 
are mainly due to the different nucleosynthetic 
prescriptions used in this paper and by D98. In particular, for stars of low and intermediate mass 
we use the metallicity-dependent yields by Van den Hoeck \& Groenewegen (1997), whereas, for stars in the same mass range,  
D98 used the yields computed by Renzini \& Voli (1981). 
For the dust production rates, in table 2 
we decided not to show any observational value since 
the only existing observational 
constraints are the ones by Jones \& Tielens (1994), which are provided without any error bar.\\
Differences in the dust accretion rates $DAR$ are due to the different present day 
gas densities and to the different nucleosynthesis prescriptions, whereas the small differences 
in the destruction rates $DDR$ are due to the different star formation laws, which cause differences 
in the predicted SNRs. Finally, in the last two lines of table 2,  we compare the C and Si dust to gas ratios as 
computed by us at the time 
when the solar system formed (i.e. at 9.5 Gyr), with the values as predicted by D98.

\begin{table*}
\begin{flushleft}
\caption[]{Comparison between some present-day output quantities (listed in the first column) predicted by means of our model for the 
solar neighbourhood (second column), 
with the same quanties as computed by D98 (third column) and with the corresponding observational values 
(when available, fourth column). $R_{\odot}=8 kpc$ is the solar distance from the Galactic center.  
References: $a$: Rana 1991. $b$: Dickey 1993. $c$: Kulkarni \& Heiles 1987. $d$: Portinari et al. 1998.}
 \begin{tabular}{l|l|l|l}
\noalign{\smallskip}
\hline
\hline
\noalign{\smallskip}
Quantity  & Our model  & D98 model & Observed\\
\hline
\noalign{\smallskip}
ISM properties  \\
\hline
$\psi(R_{\odot}, T_{0}) (M_{\odot}pc^{-2}Gyr^{-1})$ & 2.7 & 3.8 & 2-10$^a$ \\
$\sigma_{gas} (R_{\odot}, T_{0}) (M_{\odot}pc^{-2})$ & 7 & 8.8 & 7-16 $^{b,c}$ \\
$\dot{\sigma}_{infall}(R_{\odot}, T_{0}) (M_{\odot}pc^{-2} Gyr^{-1} )$ & 1 & 0.8 & 0.3-1.5$^d$ \\
\hline
Dust production  in stars\\
\hline
$DPR_{C,LIMS}(R_{\odot}, T_{0}) (M_{\odot}pc^{-2} Gyr^{-1} )$ & $4 \times 10^{-4}$ & $2.8 \times 10^{-3}$\\
$DPR_{C,SN}(R_{\odot}, T_{0}) (M_{\odot}pc^{-2} Gyr^{-1} )$ & $6 \times 10^{-4}$ & $1.6 \times 10^{-3}$ \\
$DPR_{Si,LIMS}(R_{\odot}, T_{0}) (M_{\odot}pc^{-2} Gyr^{-1} )$ & $1.4 \times 10^{-3}$ & $3.7 \times 10^{-3}$\\
$DPR_{Si,SN}(R_{\odot}, T_{0}) (M_{\odot}pc^{-2} Gyr^{-1} )$ & $8 \times 10^{-3}$ & $1.1 \times 10^{-2}$ \\
\hline
Dust accretion in the ISM\\
\hline
$DAR_{C}(R_{\odot}, T_{0}) (M_{\odot}pc^{-2} Gyr^{-1} )$ & 0.065 & 0.035 & \\
$DAR_{Si}(R_{\odot}, T_{0}) (M_{\odot}pc^{-2} Gyr^{-1} )$ & 0.23 & 0.18 & \\
\hline
Dust destruction due to SN shocks\\
\hline
$DDR_{C}(R_{\odot}, T_{0}) (M_{\odot}pc^{-2} Gyr^{-1} )$ & 0.058 & 0.033 & \\
$DDR_{Si}(R_{\odot}, T_{0}) (M_{\odot}pc^{-2} Gyr^{-1} )$ & 0.25 & 0.19 & \\
\hline
$D_C$ &  $8.5 \times 10^{-4}$ & $1.7 \times 10^{-3}$ &   \\
$D_{Si}$ &  $4.8 \times 10^{-3}$ & $7 \times 10^{-3}$ &  \\
\noalign{\smallskip}
\hline
\hline
\end{tabular}
\end{flushleft}
\label{comparison}
\end{table*}

\subsection{Dust evolution in elliptical galaxies} 

In this Section we make use of the best values for the parameters derived above
for the S.N., in order to 
extend our formalism to the class of early-type galaxies, 
implementing eq.~\ref{eq_dust} in the model by Pipino et al. (2005). 

As a first step, we do not modify the parameters related to processes such as the dust growth, 
which are mainly governed by the microphysics, and thus they should not depend on
the galactic morphology. 
However, after the onset of the galactic wind, ellipticals are basically devoid of cold gas. Therefore,  
we assume that, during this phase, no dust accretion can occur in elliptical galaxies.  

Concerning the last term in eq.~\ref{eq_dust}, now dust is allowed to escape the ISM during the galactic wind, although
no conclusion can be drawn on its fate.
Therefore, the dust loss rate in eq.~\ref{eq_dust} switches on during the galactic wind and it
has been assumed to follow the gas flow rate. In particular, we recall that
Pipino et al. (2002, 2005) successfully reproduced the amount of metals injected 
into the intra cluster medium (ICM) 
by assuming that, at each timestep, the ejected mass $\Delta \, M$ was set by the following 
condition: ${\Delta M \over M_{\rm gas}}={\Delta E \over E_{\rm th}}$,
where $\Delta E$ was the difference between the thermal energy $E_{\rm th}$ and the gas binding energy. 
Here we assume a similar condition to hold also for dust flow rate. 
Therefore, we have that the fraction of the ISM (i.e. dust plus gas)  
which can escape, scales with $\Delta E$  as:
\begin{equation}
{\Delta M \over M_{\rm ISM}}= {\Delta M_{dust} + \Delta M_{gas} \over M_{\rm dust}+M_{\rm gas}} =
{\Delta E \over E_{\rm th}}.
\label{eq_dust_esc}
\end{equation}

We are aware that the infrared emission from grains might change the cooling function (Draine 1981). 
To tackle this issue, it is necessary to study interstellar dust from a spectro-photometric point of view. 
This will be the subject of a forthcoming paper. 

\subsubsection{The dust production rates}
\begin{figure*}
\centering
\vspace{0.001cm}
\includegraphics[height=18pc,width=18pc]{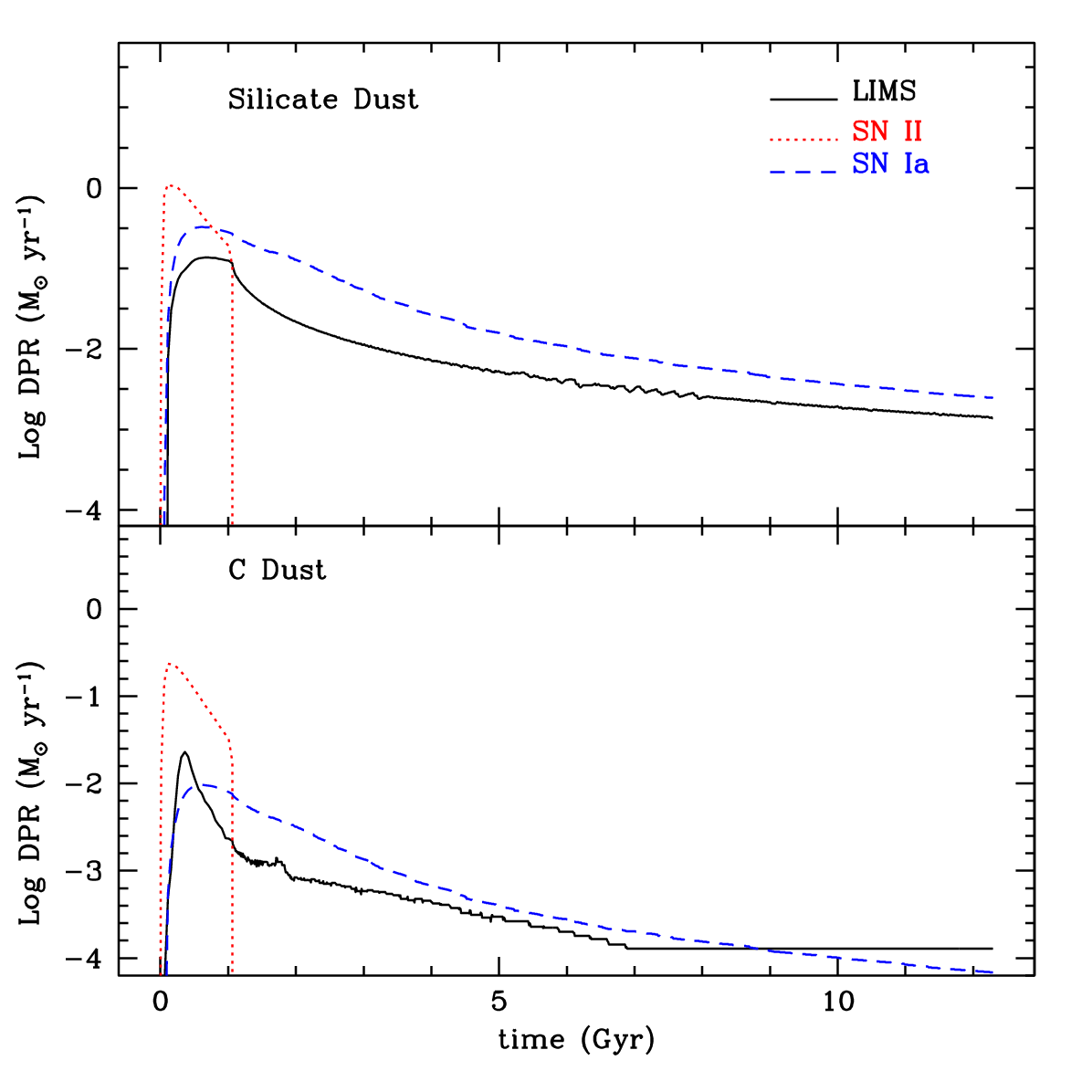}
\caption[]{
Predicted dust production rates from various sources 
for the chemical evolution model La1+Itoh (\emph{fiducial} case) for  an elliptical galaxy.  
In the lower (upper) panel, we show the results for the carbon (silicate) dust. 
Solid lines: contribution by low and intermediate mass stars (LIMS). Dotted lines: contribution 
by type II SNe. Dashed lines: contribution by type Ia SNe.}
\label{dust_ell_4}
\end{figure*}

In Figure~\ref{dust_ell_4}, we show the predicted dust production rates for the La1 model, describing an 
elliptical galaxy of stellar mass $\sim 10^{11} M_{\odot}$. During the starburst 
phase, lasting $\sim 1$ Gyr in this case, the major C and Si dust producers are type II SNe.  
After the end of the starburst, Si dust production is dominated by type Ia SNe, whereas  
C dust production is mainly due to type Ia SNe. At late times, i.e. after 10 Gyr, LIMS are 
the major C dust producers.

\subsubsection{The evolution of the dust fractions}

\begin{figure*}
\centering
\includegraphics[height=19pc,width=20pc]{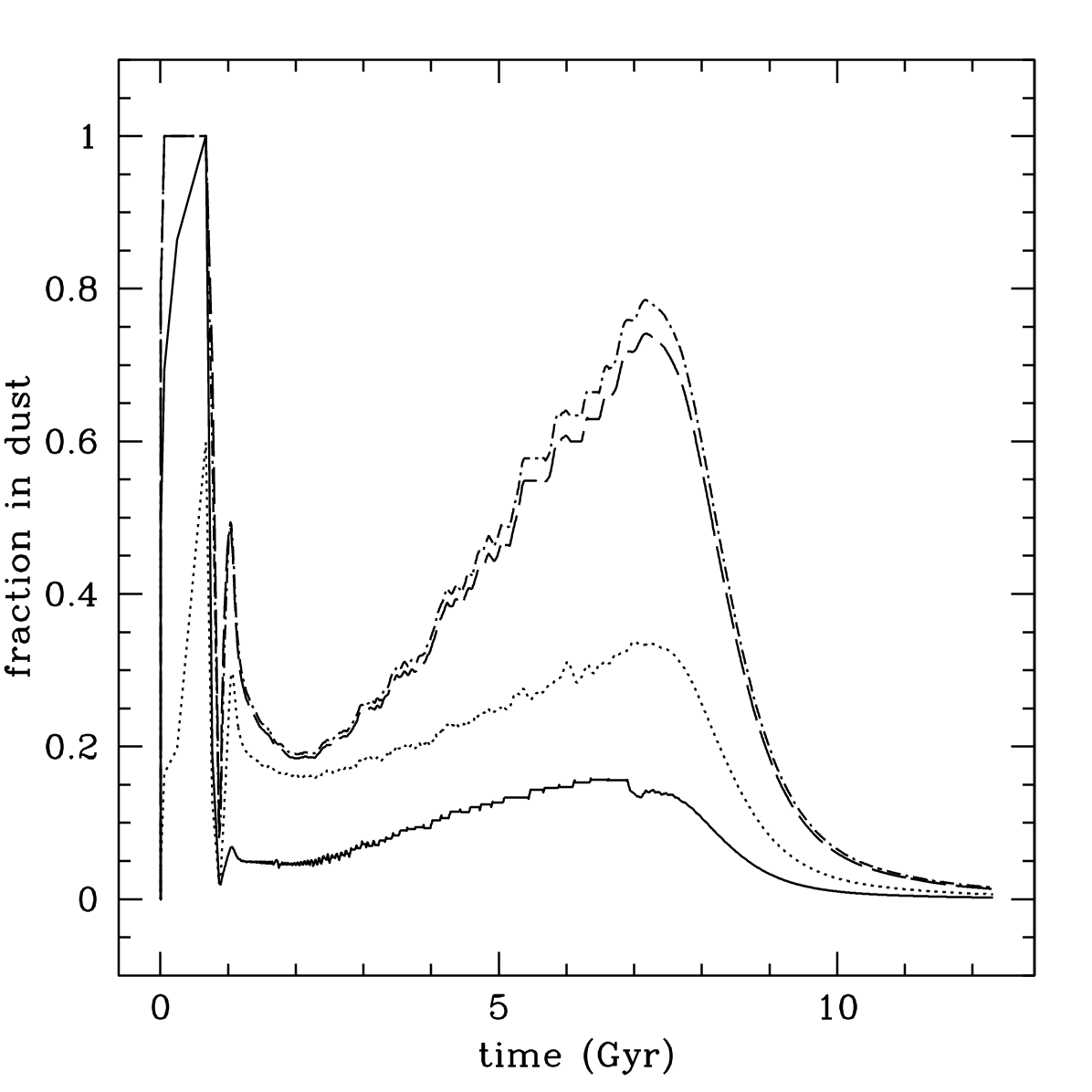}
\caption[]{
Fractions locked in dust for C (solid line), O (dotted), S (dot-dashed), Fe (dashed) for \emph{fiducial} models 
Ha1 (left panel) and La1 (right panel).
}
\label{dust_ell_3}
\end{figure*}

In Figure~\ref{dust_ell_3}, we show the predicted evolution of the dust fractions locked up in 
C, O, S and Fe for the fiducial case $\emph{La1+Itoh}$. 
In the very early phase of the evolution (i.e. $t\le$200 Myr, when
the gas is still infalling and the SFR has just reached its maximum intensity),
we predict that nearly 80\% of Si and Fe are locked into dust, whereas only 40\% of C and 10\% of O. 
This finding is in rough agreement with the depletion pattern found for spirals.
After an initial peak, the dust fractions undergo a sudden decrease
due to the high SN activity, which  leads to high destruction rates. 
The decrease in the dust fractions is then enhanced by the mass 
removal as soon as the galactic wind develops. 
After the end of the galactic wind (around 7 Gyr), the dust mass can increase again, 
before the sputtering due to the hot X-ray emitting gas can have a major role in the grain destruction. 
An interesting case is represented by the S dust fraction.  
In spirals, the evolution of the S dust fraction is similar to 
the ones of C and O, mainly because of 
the similar destruction efficiencies. In these systems, the destruction is very efficient 
because of the continuous explosions of type II SNe. 
On the other hand, dust destruction plays a minor role in ellipticals, presenting 
long periods of no 
star formation activity. In these systems, the evolution of the 
S dust fraction is close to the one of Fe, since the two elements have similar 
condensation efficiencies.

\subsubsection{The dust fractions in present-day ellipticals: local ellipticals and the Fe discrepancy}

The main answer that we expect from our analysis is whether the inclusion
of the dust treatment may help in solving the so-called iron discrepancy (Arimoto et al.\ 1997) in
X-ray spectra of the hot halos surrounding ellipticals. 
When the \emph{ASCA} satellite provided the first reliable measure of the iron
abundance in the hot ISM of ellipticals
(e.g. Awaki et al.\ 1994; Matsumoto et al.\ 1997), it
was much lower than the solar value, at odds not only with theoretical 
models for elliptical galaxies available at that time (Arimoto \& Yoshii 1987;
 Matteucci \& Tornamb\'e 1987), which predicted that their ISM should exhibit
[Fe/H]$>0$, but also with the mean metallicity of the stellar component inferred from optical spectra. 
This issue then had been partly alleviated by taking into account temperature gradients 
(Buote \& Fabian 1998; Buote 1999).
Recently, Humphrey \& Buote (2006) reported 
emission-weighted Fe abundances up to $\sim 2-3 $ solar in a sample of 28 early-type galaxies, corresponding to [Fe/H]=0.3 - 0.48. 
Nevertheless, the iron discrepancy persists when comparing the recent results by Pipino et al. (2005), 
predicting [Fe/H]$\ge 0.85 $ in the ISM of elliptical galaxies, to the most recent abundance measurements in the X-ray spectra. 

\begin{figure*}
\centering
\vspace{0.001cm}
\includegraphics[height=18pc,width=18pc]{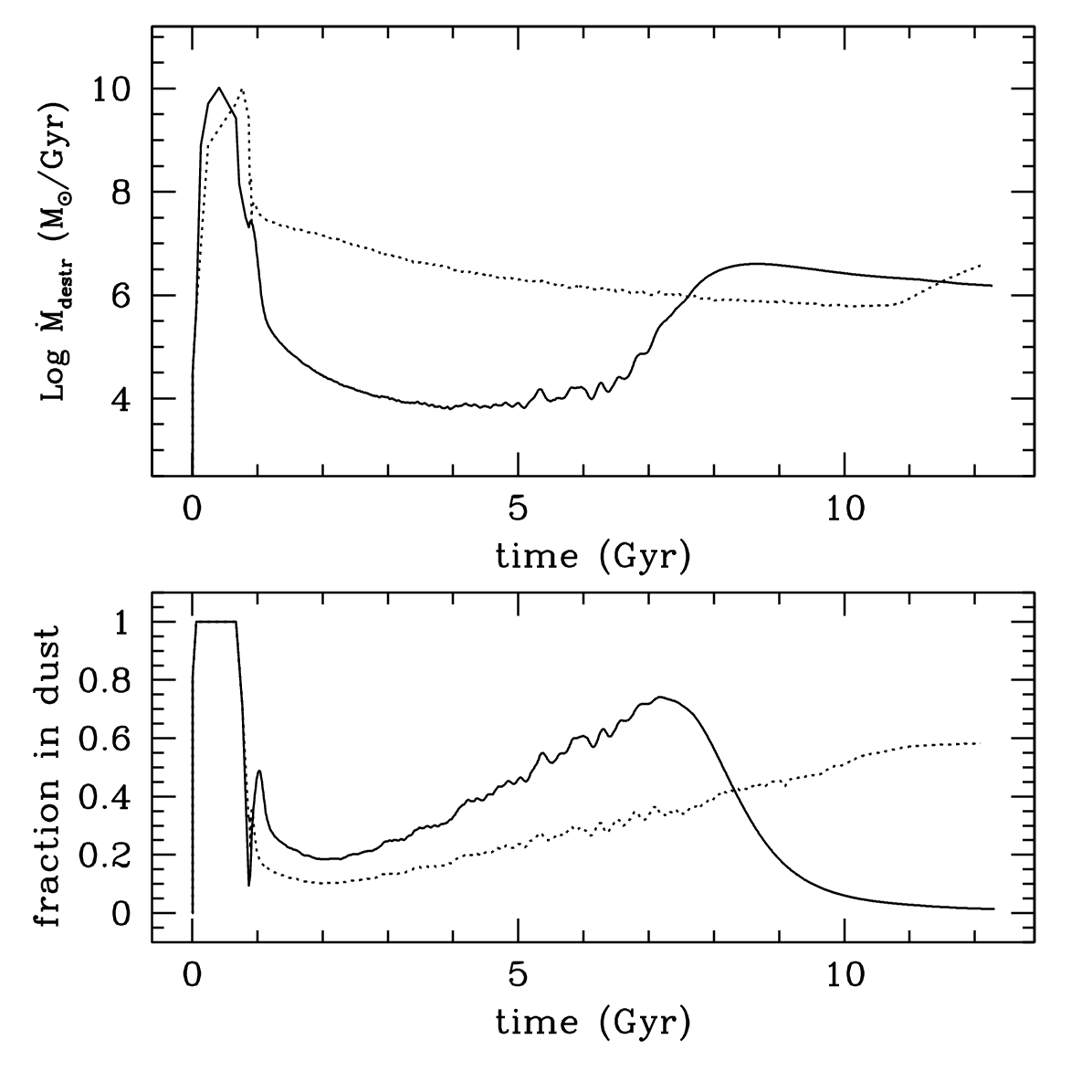}
\caption[]{Upper panel: Temporal behaviour of the Fe grain destruction rates for an elliptical galaxy (Pipino et al.(2005)'s model La1). 
             Dotted: \emph{La1+MK} case.
           Solid: \emph{La1+Itoh} case.         
	   Lower panel: Evolution of the Fe dust fraction in the ISM of the above models.}
\label{dust_ell_2}
\end{figure*}

\begin{figure*}
\centering
\vspace{0.001cm}
\includegraphics[height=19pc,width=20pc]{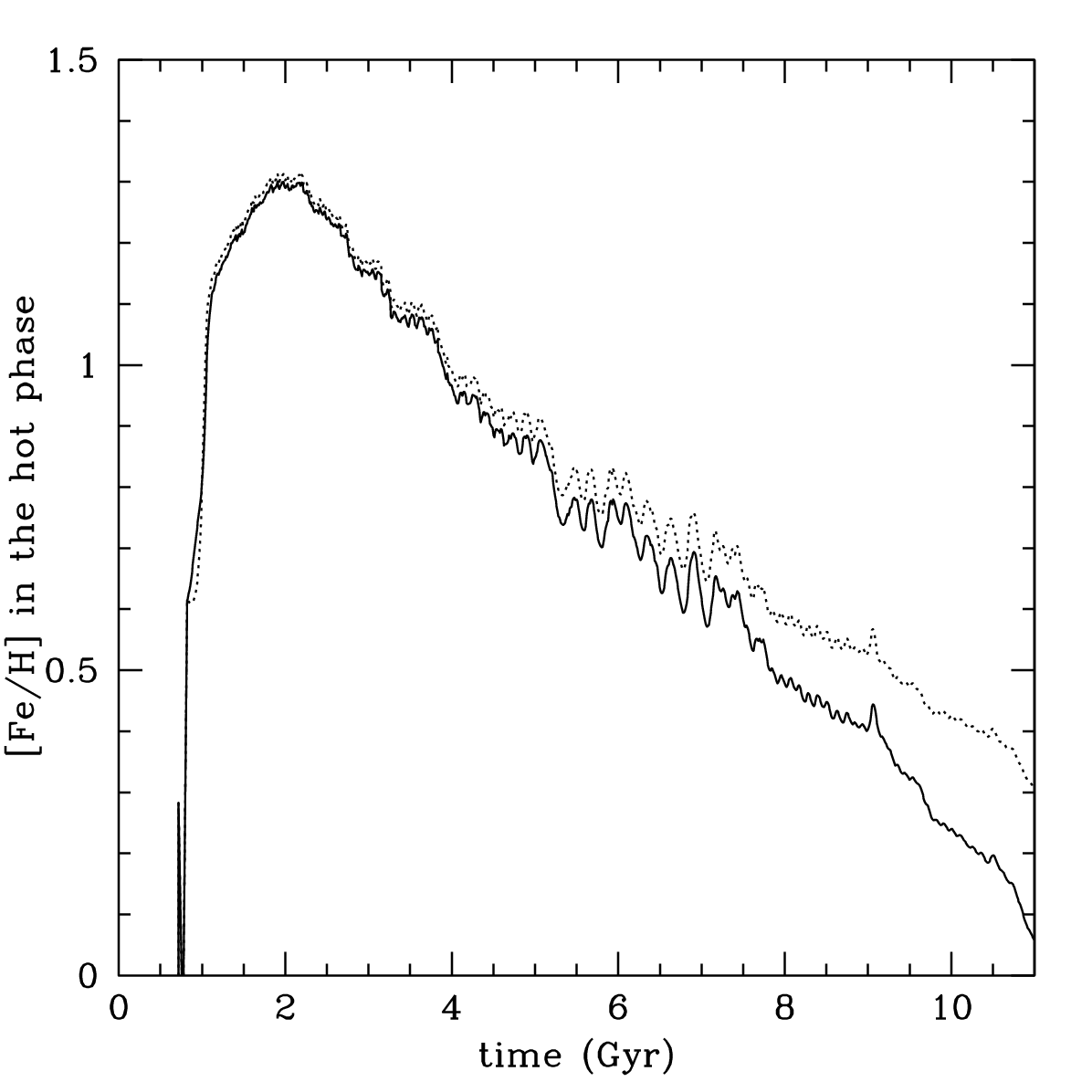}
\caption[]{Evolution of the Fe abundance in the hot phase of an elliptical galaxy for the La1 model. 
Solid lines: La1+Itoh model; Dotted Line: La1+MK model. 
}
\label{dust_Fe_disc}
\end{figure*}

\begin{figure*}
\centering
\includegraphics[height=19pc,width=20pc]{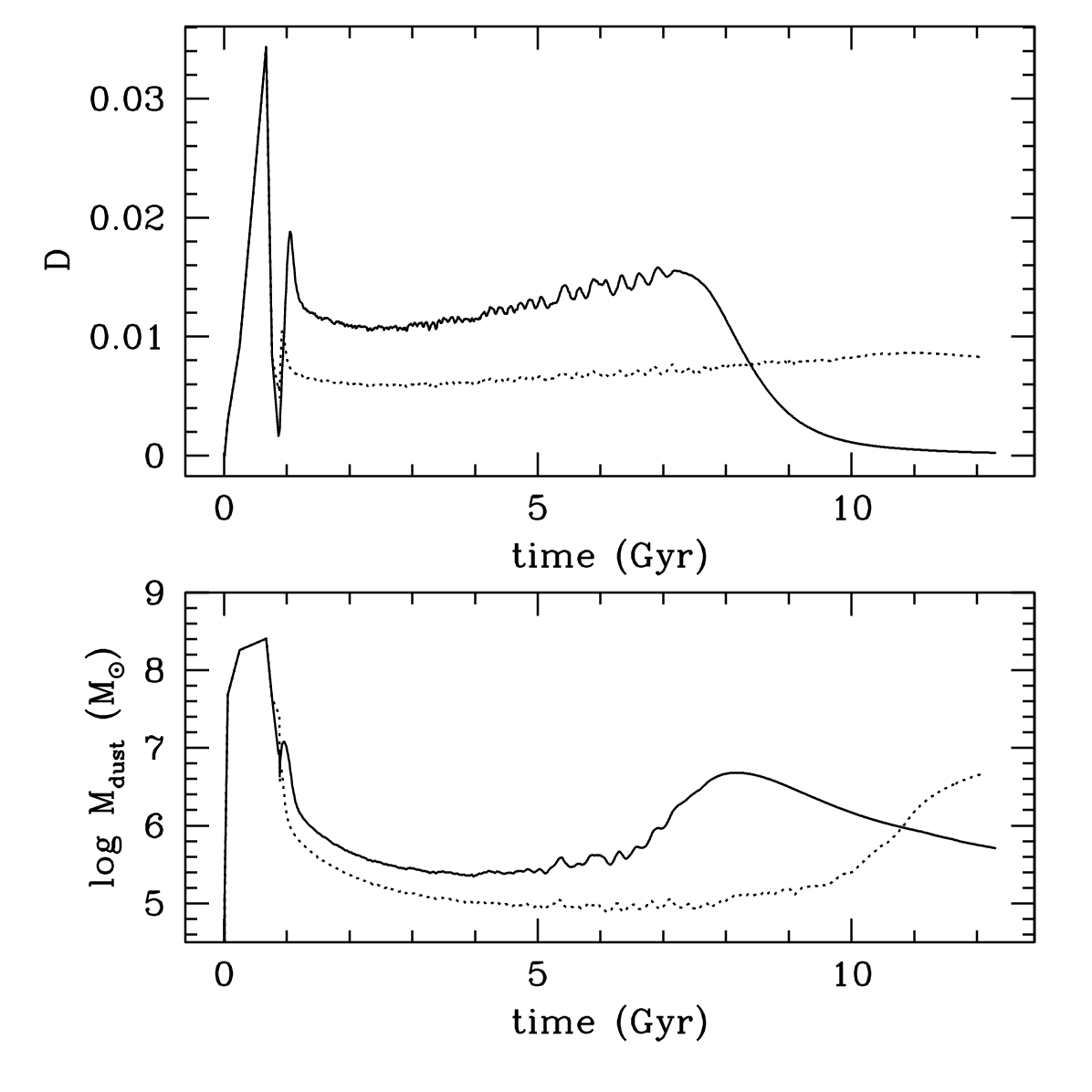} 
\caption[]{Upper panels: Dust to gas mass ratio  (\emph{D}) for Pipino et al.(2005)'s model La1. 
           Dotted: \emph{La1+MK} case.  
           Solid: \emph{La1+Itoh} case. 
           Lower panel: Evolution of the dust mass in the ISM for the above models.           
	   }
\label{dust_ell_1}
\end{figure*}

Fe condensation in dust is among the possible physical mechanisms often invoked to solve this issue (see
Arimoto et al. 1997 for a comprehensive analysis).
Recent far-infrared observations, in fact, claim that the dust mass in 
ellipticals could be $\sim 10^{6-7}M_{\odot}$,
higher than previous estimates  (Temi et al. 2004) by a factor of ten,
with $\sim 2\cdot 10^{5}M_{\odot}$ of dust residing in the very central galactic regions (Leeuw et al. 2004).
Independent recent observations carried out by means of the SCUBA camera
on local ellipticals (Vlahakis et al. 2005) seem to confirm these values, although the 
sample should be enlarged and the effects of synchrotron radiation clarified. 
Recently, several observationalal determinations of the dust masses in ellipticals have been possible 
thanks to the Spitzer Space Telescope. \\
In elliptical galaxies, Kaneda et al. (2007) derive dust masses typically of the order of $10^5-10^6 M_{\odot}$. In an early type galaxy of the 
Virgo group, Panuzzo et al. (2007) obtain a dust mass of $1.2 \times 10^6 M_{\odot}$. 
These new observational values are well reproduced by our \emph{fiducial} models, which
predict a dust mass of $\sim 1.1 \cdot 10^{6}M_{\odot}$ 
for more than 10 Gyr old giant ellipticals  ($\sim 0.8 \cdot 10^{6}M_{\odot}$
of dust for a more typical galaxy like our model La1).\\
The predicted evolution of the Fe dust destruction rates and  of the Fe fraction in dust 
for the La1 model assuming the Itoh and the McKee 
prescriptions for the destruction are presented in the upper and lower panel of Figure~\ref{dust_ell_2}, respectively. 
In Figure ~\ref{dust_Fe_disc}, we show the predicted  
evolution of the Fe abundance in the hot phase of a 
typical elliptical galaxy by adopting the Itoh (1989, solid line) and the 
Mckee (1989, dotted line) prescriptions. 
For the fiducial model, the predicted present-day iron abundance in the hot gas is
[Fe/H]=+0.6 dex (for both La1 and Ha1). These abundances are a factor of 2-5 less than the values predicted by the original Pipino et al. (2005) models La1 and Ha1, respectively.
Several factors act in determining this result: dust depletion (up to 20\% of Fe is in dust at late times),
a more prolonged wind phase with respect to the \emph{original} model Ha1 (which leads to a more efficient Fe ejection), 
different stellar yields. 
Moreover, since the Fe emission line dominates the X-ray spectrum of the hot ISM,
a reduced Fe abundance in the gas leads to a better agreement between the predicted X-ray luminosity
($2.1 \cdot 10^{42} \rm erg\, sec^{-1}$) and temperature ($\sim$ 0.7 kev) for the hot halo
of giant ellipticals with the observed values, than in Pipino et al. (2005) case.
When adopting the \emph{La1+MK} prescription (dotted lines in Figs.~\ref{dust_ell_1} and ~\ref{dust_ell_2}), 
the grain destruction is considerably reduced and up
to $\sim 4\cdot 10^{8}M_{\odot}$ of dust can survive after 12 Gyr of galactic evolution.
At this stage, the amount of Fe locked up in dust is close to 60\% (for a comparison, 
the fraction for O and C are only 25\% and 12\%, respectively), leading
to a [Fe/H]=0.15 (Ha1) - 0.3 (La1) 
dex as well as [O/Fe]=-0.12 dex, in excellent agreement with the available data
(see Pipino et al. (2005), table 4 for a compilation of the most recent ones): this might be 
the solution of the iron discrepancy. 
Contrary to other authors (Arimoto et al. 1997), we suggest that the iron discrepancy might be solved 
by taking into account the dust condensation of Fe. New and elliptical-dedicated calculations
of the sputtering by hot (i.e. $T> 10^7$K) plasmas are needed to eventually asses this issue.

\subsubsection{The dust flow into the IGM/ICM} 

The amount of dust ejected by a giant elliptical galaxy
can be obtained by integrating eq~\ref{eq_dust_esc} over time.
The amount is $\sim 3  \cdot 10^9 \, M_{\odot}$ for the \emph{fiducial} Ha1 model. 
For the model La1 (\emph{fiducial} case), instead, the ejected dust mass is 
$\sim 5 \cdot 10^8 \, M_{\odot}$. Once the dust grains are ejected into the IGM, 
no firm conclusions on their  consequent evolution can be drawn. 
For instance, we need a detailed comparison of the efficiency of dust removal by means 
of the SN-driven wind 
with the dust destruction rate due to the SN explosions. 
Moreover, a careful treatment of the dust survivability in a medium such as the hot ICM/IGM is beyond the scope of the paper.

\subsubsection{The dust fractions in young ellipticals: hints from Lyman Break and SCUBA galaxies}

Unfortunately, we do not have direct measurement of the dust composition in young ellipticals. Therefore, in this section, we will try to
constrain our models by means of the objects which are believed to be their high-redshift counterparts.

It has been argued that the so-called SCUBA sources may be ellipticals
in the process of formation (Lilly et al. 1999; Eales et al. 2000).
The inferred SFRs of $\sim 100 - 1000 M_{\odot}/yr$ are comparable to those assumed by  
our models, as well as the estimated duration of the formation process and stellar mass involved.
A useful measure of the dust mass in such high redshift objects is given by Dunne et al. (2003).
They fitted the dust mass function by means of a Schechter curve and found that the characteristic dust
mass at the break is $M_d^*\sim 4.7\cdot 10^8 \, M_{\odot}$ at redshift z=5.
Similar values can be achieved by models such as La1 during their first Gyr of evolution (we expect these
galaxies to be close to a typical present-day $L^*$ galaxy, Fig.~\ref{dust_ell_1}) and are smaller 
than the ones predicted for the Ha1 model 
which, in turn, should represent the bright end of the present-day luminosity function.

Another class of objects which can trace the evolution of low- and intermediate sized ellipticals are
Lyman Break Galaxies (LBGs). LBGs are starburst galaxies observed at high redshift ($z>2.5$)
and are identified by the colours of their far ultraviolet spectral energy distribution around the 912 \AA\ Lyman continuum discontinuity  
(Giavalisco 2002).
The interpretation of their abundance patterns suggests that LBGs may represent young spheroids, observed during their main SF episode
(Matteucci \& Pipino 2002, where we refer the reader interested in a more complete analysis), under the assumption that half of Fe was hidden in dust.
Here we briefly repeat the study of Matteucci \& Pipino (2002) by comparing
the output of the chemical evolution code with the observed abundances in MS1512-cB58, 
the brightest LBG known so far owing to its gravitationally lensed nature. 
A stellar mass of $\sim 10^{10}M_{\odot}$, an age lower than $\sim 300$ Myr and a SFR of
40$M_{\odot}yr^{-1}$ (but we can consider a range of possible values
of 20--80 $M_{\odot}yr^{-1}$ given the uncertainties affecting the 
derivation of the SFR), have been reported for MS1512-cB58 (Pettini et al. 2002).
In order to have a realistic model for such a galaxy, we extend the class
of models of Pipino et al. (2005)  down to a $M_{\rm lum}=10^{10}M_{\odot}$ spheroid, 
with a SF efficiency $\nu = 5\;\rm Gyr^{-1}$ and an infall timescale $\tau = 0.5$ Gyr.
The maximum value for the SFR is $\sim 20 M_{\odot}yr^{-1}$.
Since we focus on the very early phases of evolution, the assumptions regarding the dust destruction
in the hot halos discussed in the previous paragraphs do not affect the results.
At 130 Myr, our model predicts the following abundance ratios: [Fe/H]=-1.13, [O/H]=-0.24,
as well as [Mg/H]=-0.62 and [Si/H]=-0.64, in remarkable agreement with the values reported by Pettini et al. (2001)
and Tepliz et al. (2000). In particular, they found [Fe/H]=-1.10, [O/H]=-0.26,
as well as [Mg/H],[Si/H]$\sim$ -0.42, with quoted uncertainties of about 0.2 dex.
Pettini et al. (2002) claimed that part (roughly half) of the Fe could be locked into dust.
On the other hand,
the theoretical expectation (based on the S.N.) that also Mg and Si should be depleted, although to a lesser extent, relative to S,
was not satisfied by their data. 
An answer can be given by our model, which, in fact, predicts that
at $\sim$100 Myr the dust fractions are 50\%, 37\%, 32\% and 7\% for Fe, Mg, S and O
respectively.
Moreover, by means of our model we can \emph{a-posteriori} prove the hypothesis made
by Matteucci \& Pipino (2002) for the Fe depletion was right. The only difference
between their model and this work is in the age estimate. In particular,
we suggest here  an age of $\sim$ 100 Myr, whereas the previous estimate was close to 30 Myr.
A slightly older age can be understood in terms of the formation process.
The model used by Matteucci \& Pipino (2002) tightly followed the monolithic scenario, therefore
the SFR was maximum at the beginning.  
The present model, owing to a non-negligible infall (e-folding time of 0.5 Gyr), 
takes a longer period to assemble the same amount of stars.

For a more general comparison with LBGs, we find that the value for the dust to gas ratio $D=0.09D_{\odot}$ predicted by this last model
is in the range of values predicted by Inoue (2003). 
Moreover, concerning the quantity $D_{Z}$, we estimate a value of 0.13 (i.e. 0.2 times the S.N. value)
which is consistent with the range 0.01-1$ D_{Z, \odot}$, as predicted by Inoue (2003).
At the same age, model La1 features $D_{Z}  \sim0.7\ge D_{Z, \odot}$; this value might reflect
the fact that models such as La1 begin to be too massive to properly represent LBGs.
These results reinforce the analysis made by Matteucci \& Pipino (2002), therefore
we conclude that LBGs are very likely to be the progenitors of present-day low mass spheroids.\\

\subsection{Dust evolution in irregular galaxies} 

\subsubsection{Dust Production Rates} 
\begin{figure*}
\leftline{\includegraphics[height=18pc,width=18pc]{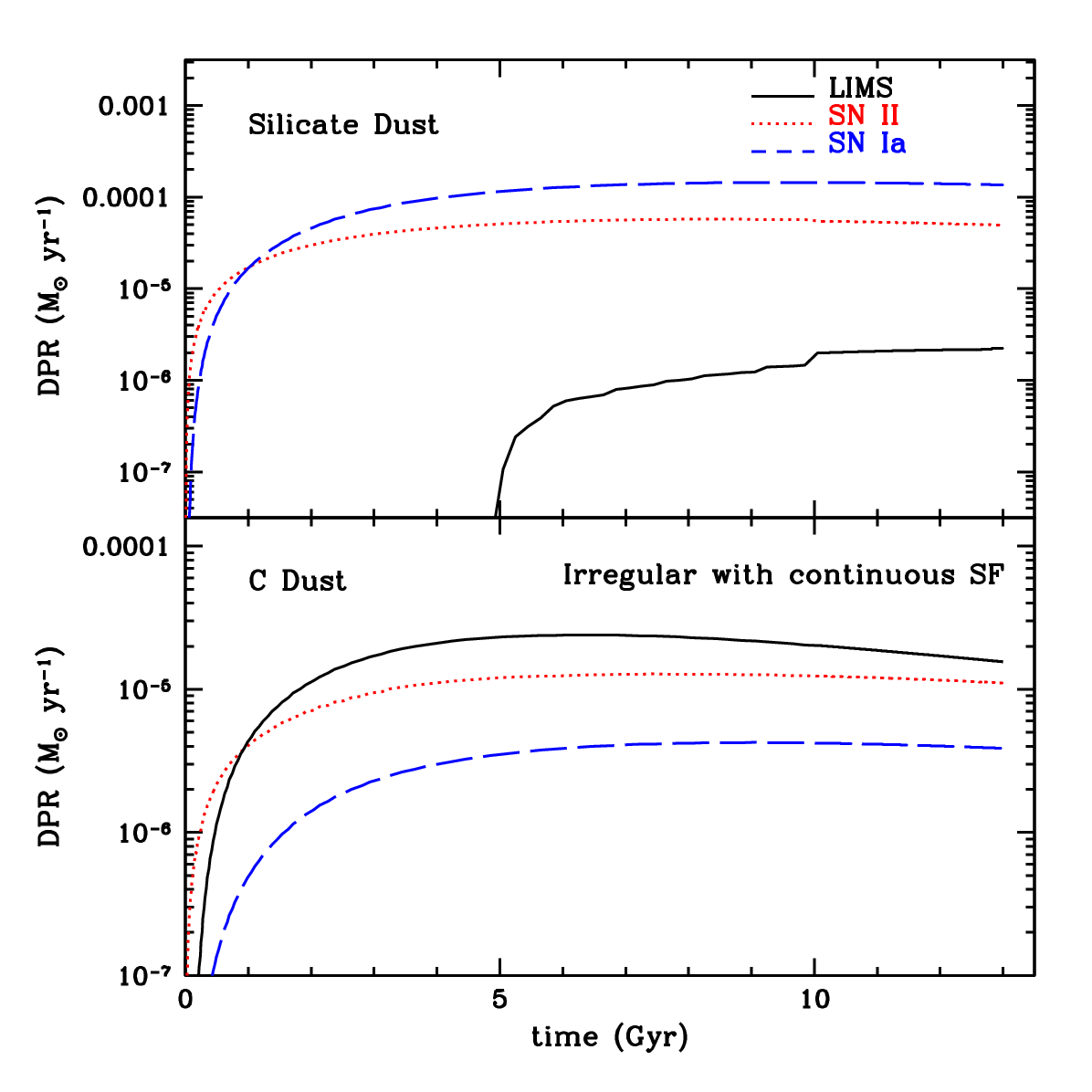}
\leftline{\includegraphics[height=18pc,width=18pc]{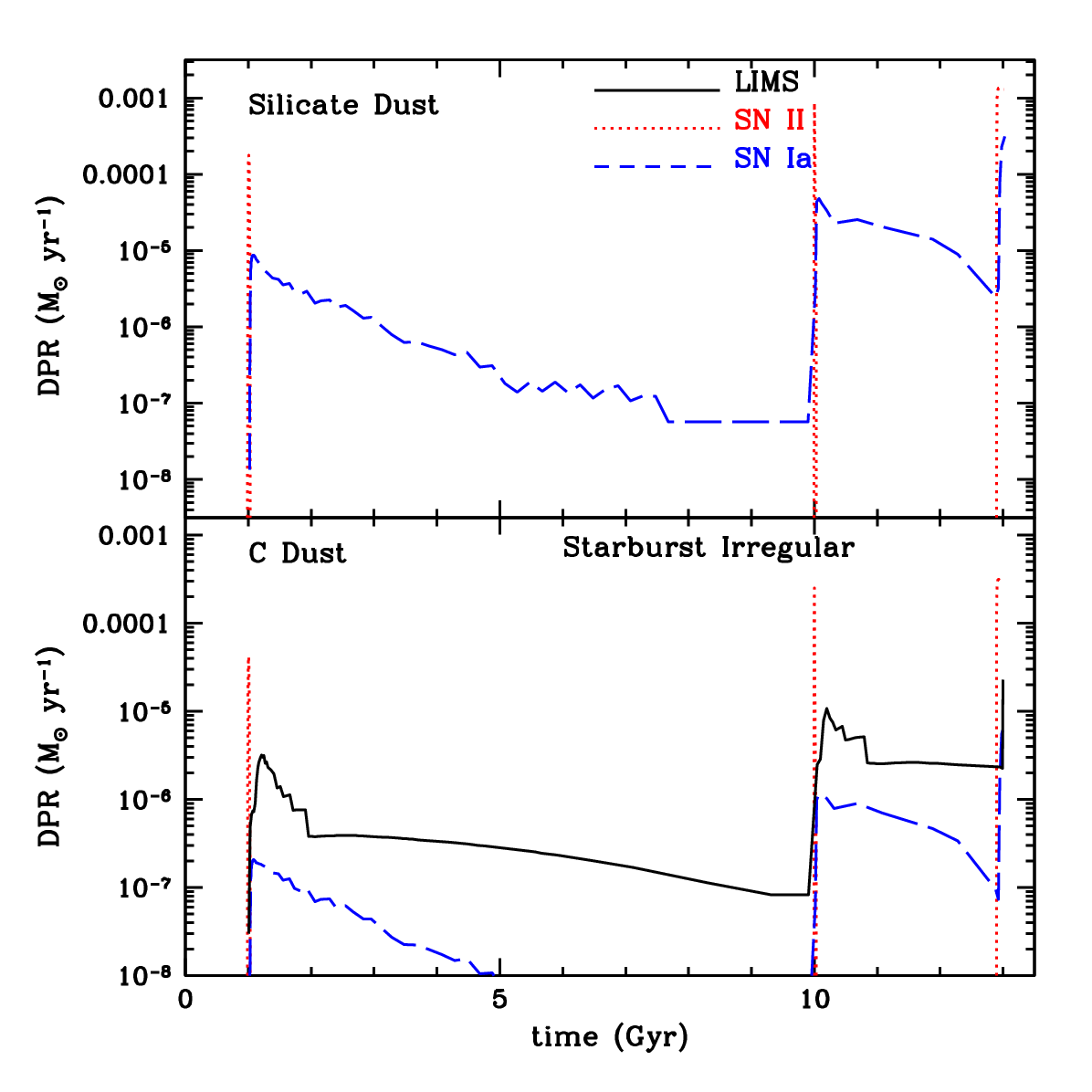}} } 
\caption[]{ Predicted dust production rates from various sources 
for a chemical evolution model of an irregular galaxy with continuous SF (left) and a starburst irregular galaxy (right).  
In the lower (upper) panels, we show the results for the carbon (silicate) dust. 
Solid lines: contribution by low and intermediate mass stars (LIMS). Dotted lines: contribution 
by type II SNe. Dashed lines: contribution by type Ia SNe.}
\label{IC+SB}
\end{figure*}

In Figure ~\ref{IC+SB}, 
we show the dust production rates for the models of an irregular galaxy with continuous SF and a starburst irregular.  
All the dust production rates calculated for the IC model have a smooth behaviour and in general present a continuous increase,  
lasting several Gyrs, and then reach a plateau. On the other hand, the production rates calculated for the starburst irregular 
reflect the shape of the SF and present a gasping or intermittent behaviour. 
An interesting difference between the dust production rates predicted for irregular galaxies and the one predicted for 
the S.N. model concerns the production of silicate dust by low and intermediate mass stars. 
In the case of the S.N. model, the Si dust production by LIMS starts immediately after the first SF episode. 
On the other hand, for the IC model Si dust production by LIMS starts at $\sim 5 $ Gyr, whereas for the starburst model it does 
not start at all. This reflects the fact that in dwarf galaxies and in low metallicity systems in general, it is rare to have 
$X_{C}/X_{O} > 1$  (see section 2.1).

\begin{figure*}
\centering
\vspace{0.001cm}
\includegraphics[height=18pc,width=18pc]{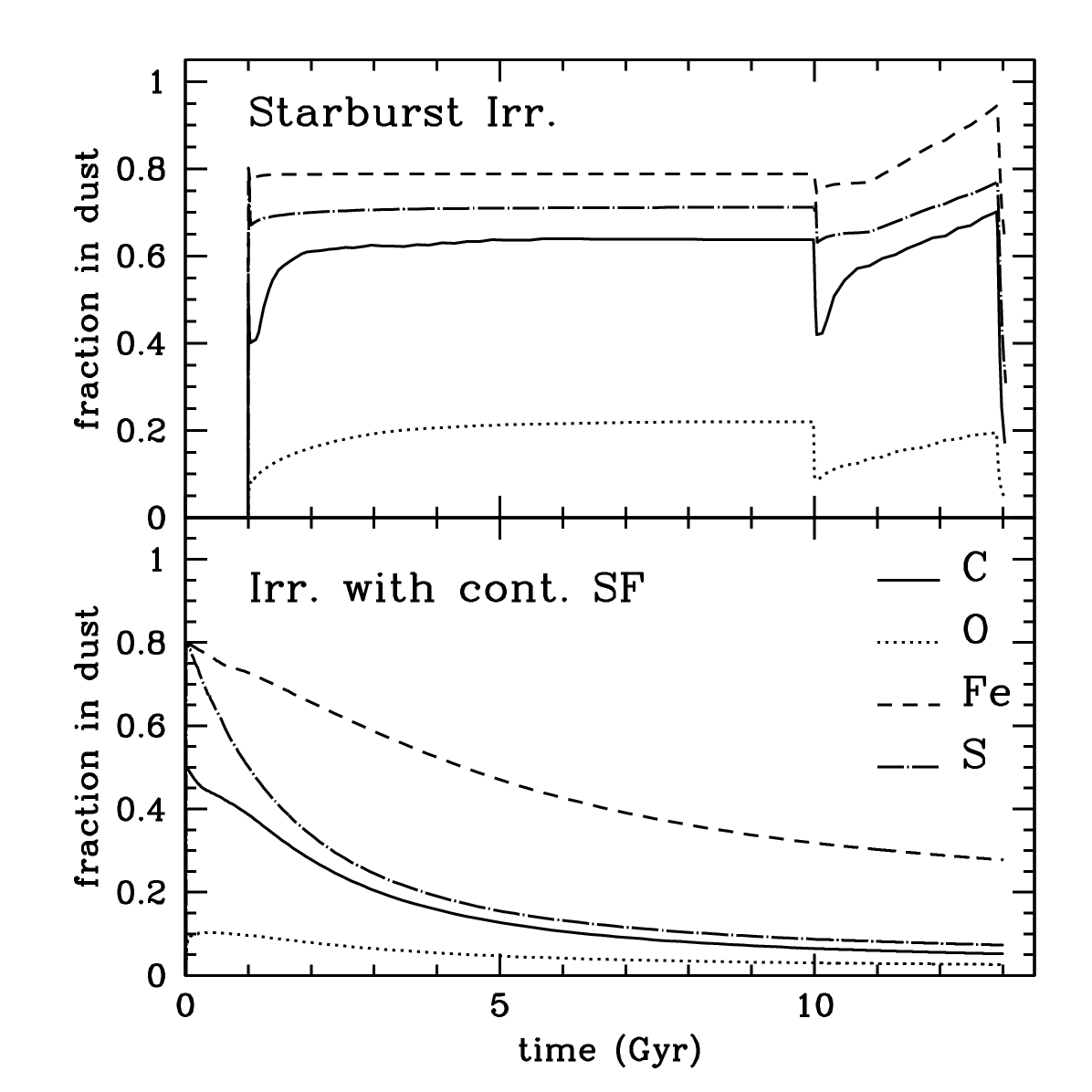}
\caption[]{Time evolution of the fractions in dust for various elements for the IC model (lower panel) 
and for the starburst model (upper panel). The Solid lines,  dotted lines, dashed lines and 
dot-dashed lines represent the evolution of the C, O, Fe and S dust fractions, respectively. 
The evolution of the dust fractions for Mg and Si is 
identical to the Fe one.  }
\label{fract_irr}
\end{figure*}

\subsubsection{The dust fractions}
\label{fract_irreg}
In Figure~\ref{fract_irr}, we show the predicted evolution of the dust fractions for the IC and the starburst irregular model. 
In the case of the IC model (Fig.~\ref{fract_irr}, lower panel), 
the dust fractions are maximum at the beginning and decrease progressively up to the present time, 
when C, O and S have dust fractions lower than $\sim 0.1 $, whereas Fe of $\sim 0.3$. These values are considerably 
lower than the ones predicted for the S.N. and for the present-day ellipticals. 
In this case, 
the main reason is the absence of the dust accretion term and the continuous SF history, 
which gives rise to continuous type II SN explosions. As seen in section 3.1.4, 
these objects are the main responsibles for dust destruction, hence play a dominant role in keeping the dust fractions 
low in irregular galaxies. \\ 
In the upper panel of Fig.~\ref{fract_irr}, we show the evolution of the dust fractions for 
the starburst irregular model. This model is characterised by long periods of 
no SF activity. During these periods, all the dust fractions tend to 
remain constant or to 
increase,  
since the only contributors to destruction are type Ia SNe. Each starburst 
corresponds to massive type II SN activity, hence to considerable dust destruction 
and consequent sharp decrease of all the dust fractions, 
which is visible at 1, 10 and 13 Gyr. 
At the present time, the predicted dust fraction for the starburst model are in 
general higher than the ones calculated for the IC model.

\begin{figure*}
\centering
\vspace{0.001cm}
\includegraphics[height=18pc,width=18pc]{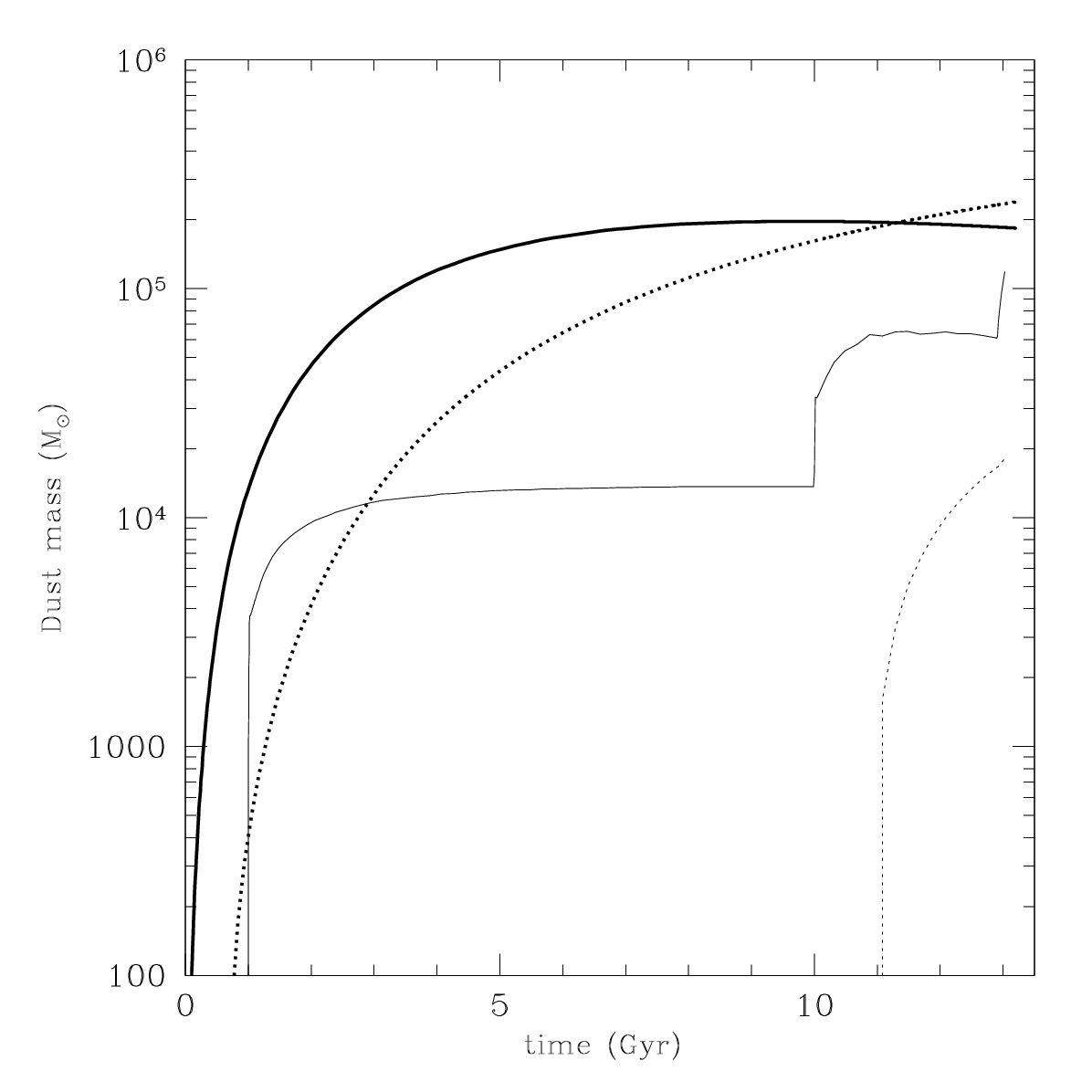}
\caption[]{Time evolution of the dust masses for the IC (thick lines) and starburst irregular (thin lines)   
models. Solid lines: dust mass 
locked up into the galaxies. Dotted lines: the dust masses ejected into the IGM. }
\label{mass}
\end{figure*}

\begin{figure*}
\leftline{\includegraphics[height=18pc,width=18pc]{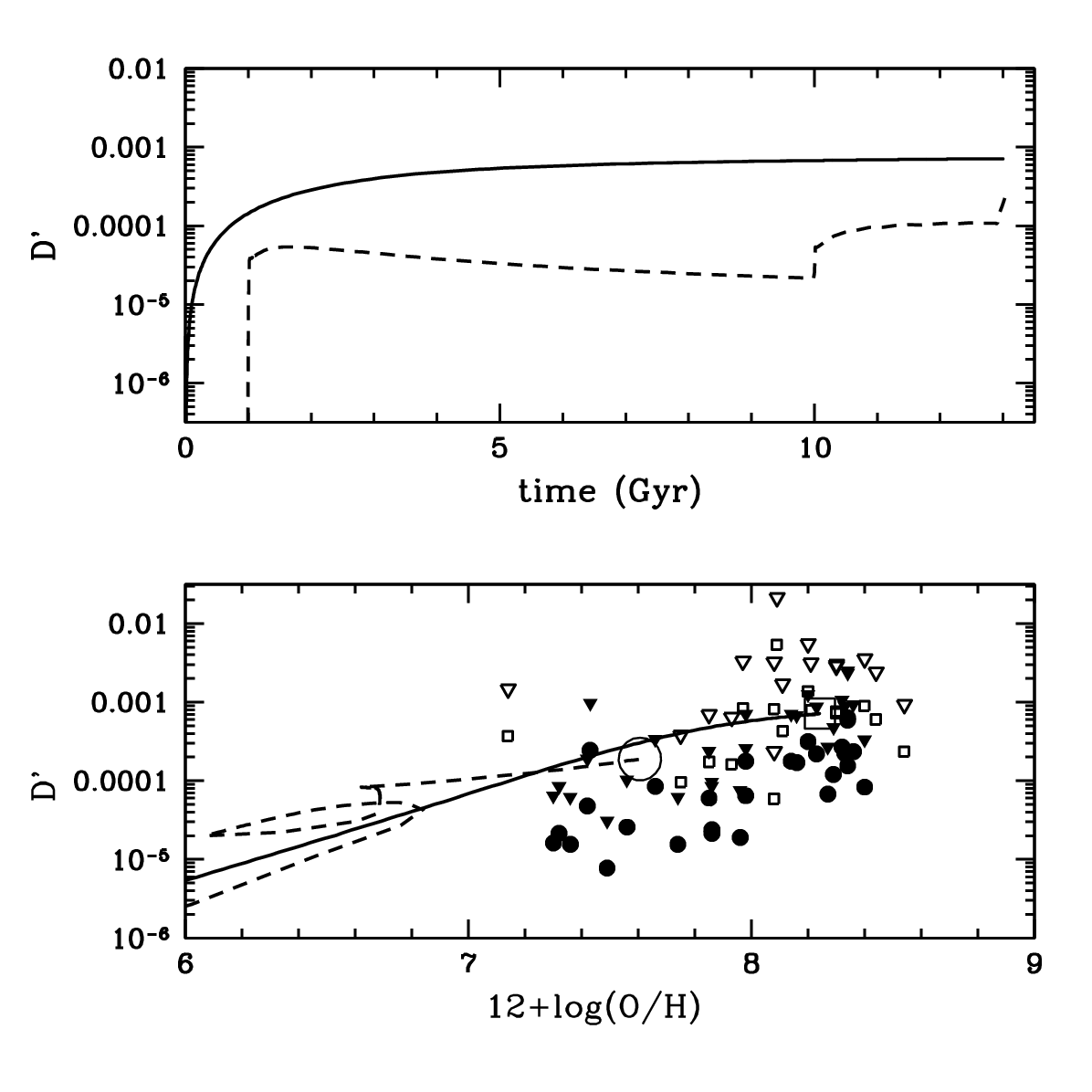}
\leftline{\includegraphics[height=18pc,width=18pc]{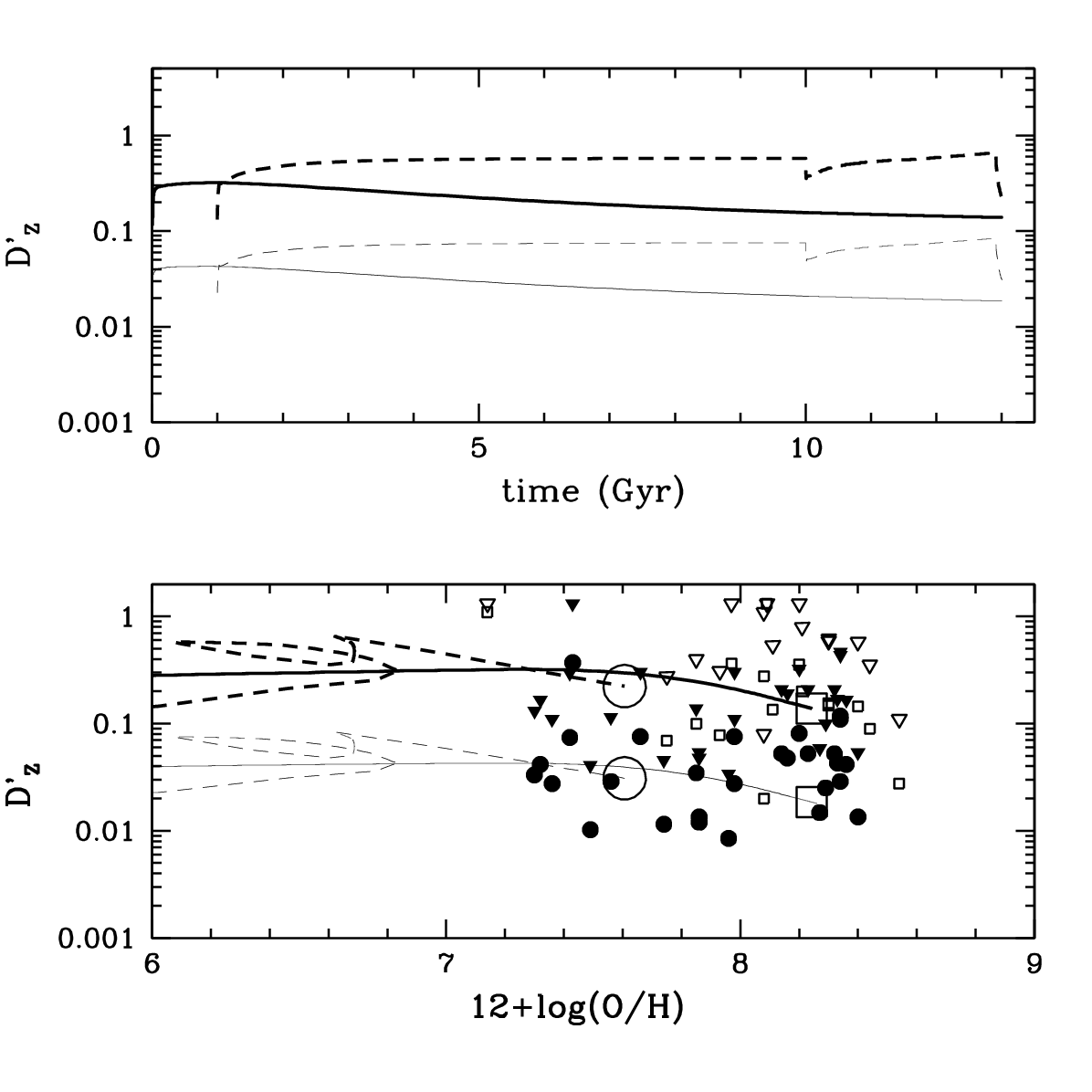}} } 
\caption[]{ \emph{Left:} Evolution of the dust to gas ratio calculated for the IC model (solid lines) 
and for the starburst model (dashed lines) as a function of time (upper panel) and metallicity (lower panel). 
In the lower panel, the solid circles and open squares are 
$D'$ values observed in local dIrr and BCD galaxies, respectively (Lisenfeld \& Ferrara 1998). 
The solid and empty inverted triangles represent upper limits 
to the observations for local dIrr and BCD galaxies, respectively, calculated by assuming an uncertainty of a factor of 4.  
\emph{Right:} Evolution of the dust to metals ratio calculated for the IC model 
(solid lines) and for the starburst model (dashed lines) as a function of time (upper panel) and metallicity (lower panel). 
The thick and thin lines represent the predicted $D'$ calculated assuming for the condensation efficiencies 
the values  suggested by D98 and   
a constant value of 0.1, respectively. 
All the symbols are as in the left lower panel. }
\label{dtg_dtm_irr}
\end{figure*}

\subsubsection{Dust masses locked up in galaxies and ejected into the IGM} 
In Figure~\ref{mass}, the thick and thin lines are the predicted evolution of the dust masses for IC and starburst 
models. The solid lines are the dust masses 
locked up into the galaxies, whereas the dotted lines are the dust masses ejected into the IGM. 
Our predictions indicate that, for an irregular galaxy of $\sim 5 \cdot 10^{8} M_{\odot}$, 
the present-day galactic dust masses are of the order of $1-2 \cdot 10^{5} M_{\odot}$, irrespective of the SF history. 
These values are compatible with the dust masses observed at the present time in dwarf irregulars (dIrr)
and blue compact galaxies (BCG), which can span from $\sim 10^{2} M_{\odot}$ up to $\sim 10^{6} M_{\odot}$ (Lisenfeld \& Ferrara 1998). 
Our values for the dust masses in local dwarfs and irregulars are compatible also with observational 
determinations achieved by means of the Spitzer Space Telescope. 
Cannon et al. (2006) derive for the local dwarf NGC 6822 a total dust mass of 8.3 $ \times 10^4 M_{\odot}$. 
For dwarf galaxies of the M81 group, Walter et al. (2007) find dust masses of $ 10^4-10^6 M_{\odot}$, in perfect agreement 
with the values we predict for dwarfs.

Larger differences are predicted for the dust masses ejected 
by IC and starburst irregular galaxies into the IGM. 
Our results indicate that, for the IC model, the galactic wind develops very early, at ages comparable to $\sim$ 1 Gyr, 
whereas for the starbust irregular the onset of the wind occurs after 10 Gyr of evolution. 
IC and starburst irregulars eject into the IGM $\sim 2 \cdot 10^{5} M_{\odot}$ and $\sim 1 \cdot 10^{4} M_{\odot}$ 
of dust, respectively. 
Calura \& Matteucci (2006a) have shown that the bulk of the IGM metal enrichment is due to giant spheroids, despite the fact that 
the number density of dwarf irregulars is larger than the one of the giant ellipticals. Since the ejected dust mass is 
proportional to the total ejected metal mass, we can conclude that ellipticals should be considered as the major contributors 
to the dust enrichment of the IGM. 

\subsubsection{Dust to gas and Dust to metals ratios} 
In the case of irregular systems, in general the observed dust to gas ratio is defined as: 
\begin{equation}
D'= \frac{M_{dust}}{M_{HI}}
\end{equation}
where $M_{dust}$ and $M_{HI}$ are the dust and $HI$ mass determined from the far infrared  
and from the radio emission, respectively (Lisenfeld \& Ferrara 1998). 
In the models, the neutral H mass is $M_{HI}=M_{gas} \cdot X_{H}$, where $M_{gas}$ is the total gas mass and 
$X_{H}$ is the H mass fraction. 
In Figure~\ref{dtg_dtm_irr}, we show the $D'$  calculated for the IC model and for the starburst model as 
a function of time (upper left panel) and as a function of the metallicity (lower left panel), 
and compared to a set of observations in local dwarf galaxies. 
At any time, the IC model is characterised by higher $D'$ values than the starburst model. \\
In the lower panel of Figure~\ref{dtg_dtm_irr}, 
as a proxy of the metallicity we use the quantity 12+log(O/H). 
The observational values used here are taken from a compilation by Lisenfeld \& Ferrara (1998) (see caption of Fig. ~\ref{dtg_dtm_irr}), 
and refer to BCD (solid circles) and dIrr (open squares) galaxies.  
The dust to gas ratio  of dwarf irregulars and BCDs as a function of the metallicity was studied by Hirashita (1999) 
and Hirashita, Tajiri \& Kamaya (2002). 
In particular, Hirashita et al. (2002) suggested that an intermittent SF history allows one to reproduce 
the observed dispersion in the relation between the dust to gas ratio and the metallicity. \\
The starburst model has a present-day metallicity lower than the IC model. 
We note that the predicted present-day values are in general higher than the majority of the observed values. 
The compilation of observations presented by Lisenfeld \& Ferrara (1998) is based on IRAS data. 
The observed determinations of the dust mass are 
likely to represent lower limits to the actual values, owing to the undetected cold dust component 
(D98, Popescu et al. 2002). 
Lisenfeld \& Ferrara (1998) estimate a total error in the dust to gas ratio to be of a factor of 4, taking into account all the possible 
factors of uncertainty (contribution from very small grains, cold dust, molecular gas and variations in the $HI$ and optical  diameter). 
By taking into account this uncertainty (solid and open inverted triangles in Fig.~\ref{dtg_dtm_irr}), 
the observed dust to gas ratios are consistent with our predictions for  both irregular types. 
The dust to metals ratio $D'_{Z}$ observed in 
dwarf irregulars can be defined as: 

\begin{equation}
D'_{Z} = \frac{M_{dust}}{M_{HI} \cdot Z} = \frac{D'}{Z}
\end{equation} 

In section 3.1.2, we have seen that the dust depletion patterns (and hence also the dust to metal ratio) 
are determined by the balance between destruction and accretion and that, in the S.N., it is impossible 
to derive constraints on the dust condensation efficiencies from the analysis of the observed depletion pattern. 
In dwarf irregulars, dust accretion is likely to play a negligible role or to be absent, hence the variation of the condensation 
efficiencies has important effects on the dust to metal ratio. 
In the right upper and lower panels of Figure~\ref{dtg_dtm_irr}, we show the predicted evolution of $D'_{Z}$ in dwarf irregular and starburst galaxies
as a function of time and metallicity, respectively, 
calculated assuming the set of 
condensation efficiencies as suggested by D98 (thick lines) and a constant value 
of $\delta^{SW}_{i}=\delta^{Ia}_{i}=\delta^{II}_{i}=0.1$ (thin lines). 
At variance with the S.N., the adoption of different condensation efficiencies has noticeable effects 
on the dust to metal ratio. In particular, the  $D'_{Z}$  values calculated by assuming the D98 condensation efficiencies 
are larger than the ones predicted by assuming a constant value of 0.1.  \\

In the lower right panel of Fig.~\ref{dtg_dtm_irr}, we show the predicted evolution of $D'_{Z}$ as a function 
of the metallicity for the IC model (solid lines) and for the starburst model (dashed lines). 
The big open circles and the big open squares are the predicted present-day values for $D'_{Z}$ 
for the starburst and IC model, respectively. 
By comparing the different predictions for $D'_{Z}$, calculated assuming two different sets of dust condensation efficiencies, 
 we can note that in principle, 
the measure of the dust to metal ratio and of the dust depletion pattern in dwarf irregulars could 
allow one to put solid constraints on the dust condensation efficiencies.\\

\renewcommand{\baselinestretch}{1.0}
\begin{table*}
\centering
\begin{tabular}{lccccccc}
\\[-2.0ex] 
\\[-2.5ex]
\hline
\multicolumn{1}{c}{Model}&\multicolumn{1}{c}{}&\multicolumn{1}{c}{DPR}&\multicolumn{1}{c}{DDR}&\multicolumn{1}{c}{DAR}&\multicolumn{1}{c}{$\sigma_{dust}$}&\multicolumn{1}{c}{$D$}&\multicolumn{1}{c}{$D_{Z}$}\\
\multicolumn{1}{c}{}&\multicolumn{1}{c}{}&\multicolumn{1}{c}{($M_{\odot} \, pc^{-2} \, yr^{-1}$)}&\multicolumn{1}{c}{($M_{\odot} \, pc^{-2} \, yr^{-1}$)}&
\multicolumn{1}{c}{($M_{\odot} \, pc^{-2} \, yr^{-1}$)}&
\multicolumn{1}{c}{(M$_{\odot}\, pc^{-2} $)}&\multicolumn{1}{c}{}&\multicolumn{1}{c}{}\\
\hline
\hline
\\[-1.0ex]
Milky Way, S.N.       &       &   0.01                     &    0.3      &   0.3         &     0.055                 &      0.008   &      0.6 \\
\hline
 Milky Way, 16 Kpc   &       &   3.8 $\times 10^{-5}$     &   1.3 $\times 10^{-4}$      &     8.7 $\times 10^{-5}$  &     0.008     &    0.0011    &    0.85  \\
\hline
\hline
\multicolumn{1}{c}{Model}&\multicolumn{1}{c}{}&\multicolumn{1}{c}{DPR}&\multicolumn{1}{c}{DDR}&\multicolumn{1}{c}{DAR}&\multicolumn{1}{c}{$M_{dust}$}&\multicolumn{1}{c}{$D$}&\multicolumn{1}{c}{$D_{Z}$}\\
\multicolumn{1}{c}{}&\multicolumn{1}{c}{}&\multicolumn{1}{c}{($M_{\odot} \, yr^{-1}$)}&\multicolumn{1}{c}{($M_{\odot} \, yr^{-1}$)}&
\multicolumn{1}{c}{($M_{\odot} \, yr^{-1}$)}&
\multicolumn{1}{c}{(M$_{\odot}$)}&\multicolumn{1}{c}{}&\multicolumn{1}{c}{}\\
\hline
\hline
Elliptical, La1  &   &  2.5 $\times 10^{-3}$  &  2.0 $\times 10^{-3}$   &    0   &  0.7 $\times 10^{6}$  &   0.0003    &    0.01         \\
\hline
Elliptical, Ha1  &   &  0.024                &    2.0 $\times 10^{-2}$  &    0   &  1.1 $\times 10^{6}$  &   0.0003    &    0.011       \\
\hline              
\hline
\multicolumn{1}{c}{Model}&\multicolumn{1}{c}{}&\multicolumn{1}{c}{DPR}&\multicolumn{1}{c}{DDR}&\multicolumn{1}{c}{DAR}&\multicolumn{1}{c}{$M_{dust}$}&\multicolumn{1}{c}{$D'$}&\multicolumn{1}{c}{$D'_{Z}$}\\
\multicolumn{1}{c}{}&\multicolumn{1}{c}{}&\multicolumn{1}{c}{($M_{\odot} \, yr^{-1}$)}&\multicolumn{1}{c}{($M_{\odot} \, yr^{-1}$)}&
\multicolumn{1}{c}{($M_{\odot} \, yr^{-1}$)}&
\multicolumn{1}{c}{(M$_{\odot}$)}&\multicolumn{1}{c}{}&\multicolumn{1}{c}{}\\
\hline              
\hline
Irregular, IC       &       &    2 $\times 10^{-4}$        &    2 $\times 10^{-4}$            &    0             &     2 $\times 10^{5}$             &   8   $\times 10^{-4}$ & 0.16\\
\hline
Irregular, burst    &       &    1.7 $\times 10^{-3}$       &  1.0 $\times 10^{-3}$           &    0             &     1.2 $\times 10^{5}$            &  2   $\times 10^{-4}$ & 0.24  \\
\hline
\hline
\end{tabular}
\caption{}
Predicted present-day dust properties for the galactic models used in this work. 
The various models  are listed in column 1. In  
columns 2, 3, 4, 5, 6 we present the predicted present day dust production rates, 
dust destruction rates, dust accretion rates, dust masses, dust to gas ratios 
and dust to metals ratios, respectively. 
\end{table*}

\section{Conclusions}
By means of chemical evolution models for galaxies of different morphological types, 
we have performed a study of the cosmic evolution of the dust properties in different environments. 
By adopting the same formalism as developed 
by D98, we have studied the evolution of the dust content of the S.N., 
confirming the main results of D98.  We have carried on a deep study of the space of the parameters used to model 
dust evolution and, thanks to the uptodate observations available in the solar vicinity, 
performed a fine tuning of the parameters.  In the following, we outline only 
the new aspects which have been investigated here and which are different than the ones considered by D98. 
We have extended our study to ellipticals and dwarf irregular galaxies, for which  
dust evolution has been calculated by means of chemical evolution models relaxing the instantaneous recycling 
approximation, namely taking into account the stellar lifetimes. 
The main results concerning the present-day dust properties for the galaxies studied here are presented in table 3.  
Our main results can be summarised as follows. \\
1) For each chemical element, we have compared the predicted present-day fractions to the ones 
observed in the Local Interstellar Cloud. This analysis has been 
useful to test the various parameters involved in our study, i.e. 
the dust condensation efficiencies $\delta^{SW}_{i}, \delta^{Ia}_{i}$, $\delta^{II}_{i}$ and the dust destruction efficiency 
$\epsilon$. 
In the most realistic physical situation, i.e. 
once both the dust destruction and accretion processes are taken into account, 
for all the elements the dust fractions are nearly independent from 
the choice of the dust condensation efficiencies. 
The main processes determining 
the gas fractions are dust accretion and dust destruction. 
We can reproduce the observed dust fractions by assuming that 
the dust destruction efficiency depends on the properties of a given chemical element. 
A physical justification for this assumption might be found in the different condensation temperatures $T_{c}$ of the elements. 
These assumptions allow us to reproduce also the gas and cosmic 
abundances observed in the Local Interstellar Cloud. \\
2) For all the elements, the evolution of the dust fractions is strongly determined by the SF history. 
At the beginning of each SF event, the dust fractions present a sharp peak, immediately followed by a very steep 
decrease, due to dust destruction by intense type II SN explosions. \\
During most of the cosmic time, i.e. from $\sim 3$ up to 10 Gyr, in the S.N. the dust fractions rise monotonically. 
The effect of the SF threshold causes rapid oscillations in the dust fractions
only  in the last 3 Gyr of evolution. \\
3)  For the dust destruction rate, we predict present-day values of $\sim 5 \times 10^{-9}$yr$^{-1}$
and  $\sim 8 \times 10^{-9}$yr$^{-1}$ for silicate and carbon dust, respectively. Similar values are predicted for the accretion rates 
for both dust types. Our values are in good agreement with the estimates by Draine (1990) and 
Tielens (1998) for a cold medium with different approaches. \\
4) For both type Ia and type II SNe, we compare the production and destruction rates 
and how these quantities vary throughout the evolution of the S.N. 
For type II SNe, 
dust destruction dominates over dust production 
throughout almost all the cosmic history. On the other hand, for type Ia SNe 
during the first Gyr of galactic evolution, Si dust production has dominated over 
dust destruction, but with negligible consequences on the total Si dust mass. \\
5) We study the evolution of the dust to gas and dust to metals ratio as a function of time and metallicity 
in the S.N. and in the outermost regions of the Galactic disc. 
The disc outskirts evolve with a SF activity less intense than the S.N., with 
a lower present-day dust to gas ratio and a higher present-day dust to metals ratio. 
For both quantities, we compare the value predicted by means of the S.N. model to the one observed in the Galaxy, 
finding a very good agreement. \\
6) In elliptical galaxies, type Ia SNe are the major dust factories in the last 10 Gyr. 
With our models, we successfully reproduce the dust masses observed in local ellipticals ($\sim 10^{6}M_{\odot}$)
by means of recent FIR and SCUBA observations. \\
7) We have shown that, by considering a reduced dust destruction in 
a hot and rarefied medium, dust is helpful in alleviating the iron discrepancy observed in the 
hot gaseous halos surrounding local ellipticals (Arimoto et al. 1997). In this medium, we predict a
Fe abundance of [Fe/H]=0.15-0.6 (depending on the adopted prescription for
dust destruction) whereas without dust Pipino et al. (2005) found [Fe/H]$\ge 0.85$.  
Moreover, the inclusion of dust improves the agreement between 
the predicted X-ray temperature (0.7 keV) and luminosity ($\sim 10^{42} \rm erg\, s^{-1}$) 
for a model of a giant
elliptical and the observations.\\ 
8) The dust masses observed in high-redshift SCUBA galaxies, the most likely progenitors 
of the local giant ellipticals, are successfully reproduced by our models, which
predict up to $\sim 10^{8-9}M_{\odot}$ of dust during the high-redshift star forming phase.
Furthermore, the dust treatment is very helpful in reproducing the abundances and the dust to gas ratios 
observed in Lyman Break Galaxies.
 In the specific case of the galaxy MS1512-cB58, we confirm the assumptions on dust made by Matteucci \& Pipino (2002) and we reinforce their 
conclusions, i.e. that LBGs are very likely to be the progenitors of the local low mass spheroids.\\
9) As for the metals, we predict that ellipticals play a major role in the dust enrichment
of the IGM/ICM during galactic winds. A minor fraction (i.e. $\sim 10^{-4}$ of the dust mass ejected by ellipticals) is ejected by Irregulars.\\
10) The two models used to study dust evolution in dwarf irregular galaxies
present very different dust production rates. For the IC model (irregular galaxy with continuous star formation), the production rates have a smooth behaviour 
throughout the whole evolution of the system. For the starburst model, the evolution of the 
production rates reflects its intermittent SF history. 
Owing to the different SF histories, the starburst model presents higher dust fractions than the IC model at any time. \\
11) The predicted present-day dust to gas ratios for irregular galaxies  
are in general higher than the majority of the observed values. 
However, the determination of the dust mass in these galaxies is affected by several sources of uncertainty.  
By assuming a factor of 4 uncertainty in the observed data, as suggested by Lisenfeld \& Ferrara (1998), 
the observed dust to gas ratios are consistent with our predictions. \\
12) In dwarf irregulars, dust accretion is likely to play a negligible role or to be absent, hence the variation of the condensation 
efficiencies has important effects on the dust to metal ratio. 
In principle, a precise determination of the dust to metal ratios and of the depletion pattern 
in dwarf irregular  galaxies could be helpful to put solid constraints on the dust condensation efficiencies. \\
13) Finally, it is interesting to compare our results for the present-day dust masses predicted for ellipticals and irregulars with 
the recent observational determinations, possible thanks to the Spitzer Space Telescope. \\
In local ellipticals, Kaneda et al. (2007) and Panuzzo et al. (2007) 
derive dust masses typically of the order of $10^5-10^6 M_{\odot}$. 
The values computed  by means of our 
fiducial model are compatible with the observationally derived values quoted above (see table 3). \\
Also our predictions for the dust masses of dwarfs/irregular galaxies are in agreement with the Spitzer results. 
In fact, for local dwarf galaxies, dust masses of $ 10^4-10^6 M_{\odot}$ have been detected (Cannon et al. 2006,  Walter et al. 2007), 
in perfect agreement 
with the values we predict for present day dwarf galaxies, i.e. $1-2 \times 10^5 M_{\odot}$.

\begin{acknowledgements}
We wish to thank Giovanni Vladilo, John Danziger and Roberto Maiolino for many helpful discussions and suggestions. 
\end{acknowledgements}

%
%

\end{document}